\documentclass[twocolumn,superscriptaddress,showpacs,preprintnumbers,amsmath,amssymb,prb]{revtex4}
\usepackage{graphicx}
\usepackage{dcolumn}
\usepackage{bm}
\usepackage{mathrsfs}
\usepackage{wasysym}

\begin{document}

\title{Strong electron-photon coupling in one-dimensional quantum dot chain:\\ Rabi waves and Rabi wavepackets}


\author{G. Ya. Slepyan}%
\author{Y. D. Yerchak}
\email{Jarchak@gmail.com}

\affiliation{%
Institute for Nuclear Problems, Belarus State University,
Bobruiskaya 11, 220050 Minsk, Belarus
}%

\author{A. Hoffmann}
\affiliation{Institut f\"{u}r Festk\"{o}rperphysik, Technische
Universit\"{a}t Berlin, Hardenbergstrasse 36, 10623 Berlin, Germany}

\author{F. G. Bass}
\affiliation{Department of Physics, Bar-Ilan University, 52900 Ramat-Gan, Israel}


\begin{abstract}
We predict and theoretically investigate the new coherent effect of nonlinear quantum optics -- spatial propagation of Rabi oscillations (Rabi waves) in one-dimensional quantum dot (QD) chain.  QD-chain is modeled by the set of two-level quantum systems with tunnel coupling between neighboring QDs. The space propagation of Rabi waves in the form of traveling waves and wave packets is considered. It is shown, that traveling Rabi waves are quantum states of  QD-chain dressed by radiation. The dispersion characteristics of traveling Rabi waves are investigated and their dependence on average number of photons in wave is demonstrated. The propagation of Rabi wave packets is accompanied by the transfer of the inversion and quantum correlations along the QD-chain and by the transformation of quantum light statistics. The conditions of experimental observability are analyzed. The effect can find practical use in quantum computing and quantum informatics.
\end{abstract}

\maketitle

\section{Introduction}
\label{sec:Intro}

Rabi oscillations are periodical transitions of a two-level quantum system between its stationary states under the action of an oscillatory driving field, see e.g. \cite{Scully, Cohen-Tannoudji_1998}. The phenomenon was theoretically predicted by Rabi on nuclear spins in radio-frequency magnetic field \cite{Rabi_1937} and was firstly observed by Torrey \cite{Torrey_1949}. Afterwards, Rabi oscillations were discovered in various physical systems, such as electromagnetically driven atoms\cite{Hocker_Tang_1968} (including the case of Rydberg atomic states \cite{Johnson_08}), semiconductor quantum dots (QDs) \cite{Kamada_2001}, and different types of solid-state qubits (superconducting charge qubits based on Josephson junctions \cite{ Blais_04, Gambetta_06, Blais_07}, spin-qubits \cite{Burkard_06}, semiconductor charge-qubits \cite{Barret_03}).
In real physical systems the ideal picture \cite{Scully} of Rabi effect can be essentially modified by additional features, such as the time-domain modulation of the field-matter
coupling constant \cite{Law_96,yang_04}, the phonon-induced
dephasing \cite{forstner_03} and the local-field effects
\cite{Slepyan_04, Paspalakis_06, Slepyan_07}.
New phenomena appear in Rabi oscillators with broken inversion symmetry \cite{Kibis_09} and in systems of two coupled Rabi oscillators
\cite{Unold_05,Gea-Banacloche_06,Salen_08,
Huges_05,Danckwerts_06,Ho_Trung_Dung_02,Tsukanov_06}.

In spatially extensive samples with a large number of
oscillators the propagation effects come into play. As a result, the mechanism responsible for Rabi oscillations causes also a number of nonstationary coherent optical phenomena, such as optical nutation, photon echo, self-induced transparency, etc
\cite{Shen_84}.  In
low-dimensional systems the propagation effects also take place. For example, the
numerical modeling of the coherent intersubband Rabi oscillations
in a sample comprising 80 AlGaAs/GaAs quantum wells
\cite{Waldmueller_06} shows that the population dynamics 
depends on the quantum well position in the series. This result
demonstrates strong radiative coupling between wells and, more
generally, significant difference in the picture of Rabi effect  for
single and multiple oscillators. Another aspects of such a difference, namely,  the effects of quantum interference and  correlations between photons in multiatom fluorescence, are demonstrated  in \cite{Kien_Hakuta_08, Tsoi_Law_08} (on the example of atomic chains in nanofibers). From practical point of view, the effect of Rabi oscillations is a key ingredient for realization of binary logic and optical control in quantum informatics and quantum computing. 

The theoretical analysis of Rabi oscillations is highly diversified both in form and content. The common feature is the impossibility of consideration of electromagnetic field influence as a small perturbation.  Different ways of description are used in the analysis of the model problem of Rabi oscillations in a single two-level atom \cite{Scully}. One such a way, the probability amplitude method, consists in solving of the Schr\"odinger equation for wavefunction $\left| \Psi \right\rangle$, which is the superposition of various atom-photon states. The second one is the Heisenberg operator method which is based on the analysis of the photonic and atomic operators time evolution. And finally, the third way is the unitary time-evolution operator technique. As is demonstrated in \cite{Scully}, all these methods lead to identical solutions; the choice of the concrete one is  determined by the convenience considerations.

Taking into account the quantum nature of the electromagnetic field  has the special significance in the analysis of Rabi oscillations in complex systems. Two different cases can be marked out. The first could be called quasiclassical one \cite{Shen_84}. In this case external field has classical nature. However, the true field   is supposed to be concordant with  the quantum motions of the particles. So, the field should contain the contribution of the induced polarization, which has a quantum nature. Another case is really quantum and takes into consideration the photon structure of the electromagnetic field \cite{Scully}. During the process of energy-level transitions in the atom the photon structure of the field is also transformed, therefore the atom-photon dynamics should be considered self-consistently.

In this paper we build for the first time a theoretical model of a distributed system of coupled Rabi oscillators and predict a new physical effect: propagation of Rabi oscillations in space in the form of traveling
waves and wave packets. Quantum oscillators in the system that exhibit Rabi-wave propagation interact strongly enough to stare and exchange the e-h pair excitation between light emission and light absorption acts. Often in quantum optics an "all-matter" picture is employed, where the dynamics of electromagnetic field is integrated out, for example in the optical Bloch equations \cite{Allen_Eberly_75, Loudon_83}. In \cite{Wubs_04, Sorensen_08} the light-matter interaction is treated in an "all-light" picture (Lippmann-Schwinger equation approach).  

In our model the quantum nature of light is also fully accounted. As the theoretical approach the \textit{probability amplitude method}, generalized for the case of 1D-chain, is convenient. 

The paper is organized as follows. In Sec.\ref{sec:theory} the model is formulated and equations of electron-photon dynamics both for discrete chain and continuous limit are obtained.  In Sec.\ref{sec:trav_w} the plane Rabi waves are considered (their dispersion equations, eigenmode structure, dispersion curves properties). Sec.\ref{sec:wave_pack} is devoted to the Rabi wavepackets investigation, their spatial-temporal dynamics is analyzed, experimental observability of Rabi waves is discussed. In Sec.\ref{sec:class_lim} the Rabi waves in the case of classic light are examined and distinctive features of semiclassical  consideration are treated of. In Sec.\ref{sec:local_field} the role of the local-field effects and conditions of this effects negligibility are discussed.  In Sec.\ref{sec:correlators} the space-time structure  of electron-electron and electron-photon correlators  is considered. The main results of the work are formulated in Sec.\ref{sec:conclusion}.

\section{Model}

\label{sec:theory}

\subsection{Hamiltonian}

\begin{figure}[htb]
\includegraphics[width=0.45\textwidth] {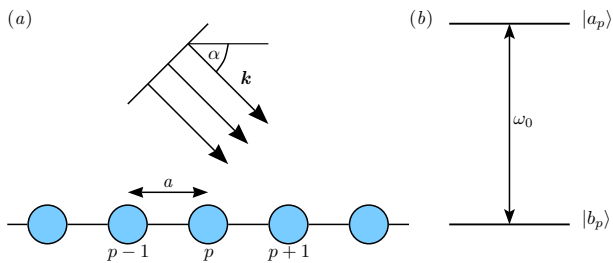}
\caption{\label{QDchain} Schematic picture of the QD-chain interacting with single-mode electromagnetic field (a) and the  energy levels diagram for $p$-th QD (b).}
\end{figure}
Consider an infinite periodical one-dimensional chain of identical QDs, containing
an electron in the size-quantized conduction band. It is assumed that in each  $p$-th QD there are at least two one-electron orbital
states, ground $\left|b_p\right\rangle$ and excited $\left|a_p\right\rangle$, with transition frequency $\omega_0$. 
Neighboring QDs are coupled via electron tunneling \cite{Tsukanov_06}, i.e. the electron can go from state $\left|a_p\right\rangle$ to state $\left|a_{p\pm1}\right\rangle$ and from $\left|b_p\right\rangle$ to state $\left|b_{p\pm1}\right\rangle$. Transitions between ground state and excited state belonging to different QDs are neglected: $\left\langle a_{p\pm1}\right.\left|b_{p}\right\rangle \approx 0$. Let the QD chain be exposed to a plane wave of quantum light, which is incident on the chain at an angle $\alpha$, $\bm{\hat {E}}(x)=  \bm{\mathcal{E}}(\hat {a}e^{ikx}+ \hat {a}^+e^{-ikx})$, where $\hat {a}$, $\hat {a}^+$ are photon creation and annihilation operators, $ \bm{\mathcal{E}}=\sqrt{2\pi\hbar\omega/V_0} \bm{e}$, $V_0$ is the normalizing volume, $\bm{e}$ is the unit polarization vector, wavenumber $k = \omega \cos \alpha/c$. The dependence of the field on transverse coordinates is negligibly small because of  electrical smallness of QDs. The case of oblique propagation is of interest due to the possibility of wavenumber variation with variation of incidence angle $\alpha$ at constant frequency $\omega$.  Then the Hamiltonian of the system in the rotating wave approximation \cite{Scully} reads 
\begin{equation} 
\label{hamiltonian}
\hat {H}=\hat {H}_d +\hat {H}_f+\hat {H}_{df}+\hat {H}_T+\Delta \hat {H},
\end{equation}
where $\hat {H}_d =(\hbar \omega _0/2) \sum_p {\hat {\sigma }_{zp}}$ is Hamiltonian of QD-chain in the absence of electron tunneling and QD-field interaction,  $\hat {\sigma}_{zp} = \left|a_p\right\rangle  \left\langle a_{p} \right|-\left|b_p\right\rangle  \left\langle b_{p} \right|$,  $\hat {H}_f =\hbar \omega \hat {a}^+\hat {a}$ is the Hamiltonian of the free electromagnetic field\cite{ref01}. The component
\begin{equation}
\label{h_df}
\hat {H}_{df} =\hbar g \sum\limits_p {(\hat {\sigma }_p^+\hat {a}e^{ikpa}+ \hat {\sigma }_p^-\hat {a}^+e^{-ikpa})}
\end{equation}
describes QD-field interaction, where $g = -\bm{\mu \mathcal{E}} /\hbar$ is interaction constant, $\bm{\mu}$ is the QD dipole moment. Note that all transition dipole moments in the chain are assumed to be real values and to have fixed orientation. The operators $\hat{\sigma }_p^{+}= \left|a_p\right\rangle  \left\langle b_{p} \right|$ and $\hat{\sigma }_p^{-}= \left|b_p\right\rangle  \left\langle a_{p} \right|$ are the transition operators from ground to excited states and vice versa for $p$-th QD.
The term
\begin{equation}
\begin{split}
\label{h_t}
\hat {H}_T=-\hbar \xi_1\sum\limits_p \left(\left|a_p \right\rangle \left\langle a_{p+1} \right|+\left|a_p \right\rangle \left\langle a_{p-1}  \right| \right)\\
-\hbar \xi_2\sum\limits_p \left( \left| b_p \right\rangle \left\langle b_{p+1}\right|+ \left| b_p \right\rangle \left\langle b_{p-1} \right|\right)
\end{split}
\end{equation}
takes into account interdot electron tunneling;  $\xi_{1,2}$ are the electron tunneling frequencies for the excited ($\xi_1$) and ground ($\xi_2$) states of the QDs. The term $\Delta \hat {H}$ corresponds to the local-field effects
originated from the dipole-dipole electron-hole intradot interaction \cite{Slepyan_04,Paspalakis_06,Slepyan_07}; in the mean-field
approximation this term is given by
\begin{equation}
\label{Deltah}
\Delta \hat{ H}=\frac{4\pi}{V}\bm{\mu}(\tilde{\underline{N}}
\bm{\mu})\sum\limits_p(\hat{\sigma}^{-}_p\langle\hat{\sigma}^{+}_p
\rangle+\hat{\sigma}^{+}_p\langle\hat{\sigma}_p^{-} \rangle).
\end{equation}
Here  $\tilde{\underline{N}}$ is the depolarization tensor (see Eq. (18) in Ref. \cite{Slepyan_07}), $V$ is the volume of QD.

\subsection{Equations of motion}
\label{sec:mot_eq}
The state vector of the "QD-chain+light" system may be represented in terms of the eigenstates of isolated QDs as
\begin{equation}
\label{state_vector}
\left| {\Psi (t)} \right\rangle =\sum\limits_n \sum\limits_p \left(A_{p,n}(t) \left|a_p,n \right\rangle +B_{p,n}(t) \left| b_p,n \right\rangle \right) .
\end{equation}
Here $\left| b_p,n \right\rangle = \left| b_p \right\rangle\otimes\left|n \right\rangle$, $\left| a_p,n \right\rangle = \left| a_p \right\rangle\otimes\left|n \right\rangle$, where  $\left|n \right\rangle$ is the light Fock state with $n$  photons, $B_{p,n}$, $A_{p,n}$ are the probability amplitudes.
In the interaction picture the evolution of the system is described by the nonstationary Schr\"odinger equation $i \hbar\partial_t\left| \Psi \right\rangle = \hat {V}\left| \Psi\right\rangle$, where the interaction Hamiltonian has the form  $\hat {V}=exp(i\hat {H}_f t/\hbar)(\hat {H}_d +\hat {H}_{df}+\hat {H}_T+\Delta \hat {H})exp(-i\hat {H}_f t/\hbar)$. This Schr\"odinger equation  leads to the following equations for the probability amplitudes
\begin{align}
\label{system_diskr_a}
 &\frac{\partial A_{p,n}}{\partial t}=-\frac{i\omega_0}{2}A_{p,n}+i\xi_1\left(A_{p-1,n}+A_{p+1,n}\right)\\
\nonumber&-i g\sqrt{n+1}B_{p,n+1} e^{i(kpa-\omega t)} -i\Delta\omega B_{p,n}\sum\limits_m A_{p,m}B_{p,m}^*,   \\ \rule{0in}{4ex}
\label{system_diskr_b}
\nonumber&\frac{\partial B_{p,n+1}}{\partial t}=\frac{i\omega_0}{2}B_{p,n+1}+i\xi_2\left(B_{p-1,n+1}+B_{p+1,n+1}\right)\\
&-i g\sqrt{n+1}A_{p,n} e^{-i(kpa-\omega t)} -i\Delta\omega A_{p,n+1}\sum\limits_m A_{p,m}^*B_{p,m},
\raisetag{35pt}
\end{align}
where
\begin{equation}
\label{dep_shift}
\Delta\omega=\frac{4\pi}{\hbar{}V}{\bm \mu}(\widetilde{{
\underline N}}{\bm \mu})
\end{equation}
 is the local-field induced depolarization shift \cite{Slepyan_07}. Obtaining this equations we have taken into account that the interaction \eqref{h_df} can cause the transitions between the states $\left| a_p,n \right\rangle$, $\left| b_p,n+1 \right\rangle$  only. As it is seen from Eqs.\eqref{system_diskr_a}--\eqref{system_diskr_b}, two competitive mechanisms
manifest themselves in the light - QD-chain coupling: the
local-field induced nonlinearity and the dispersion spreading due to
the tunneling.

To this point we have taken no account of the decay processes inside the QDs. To take this processes into account one should allow for processes of interaction of e-h pair in QD with phonon bath. Such interaction can be described by conception of quantum trajectories \cite{Plenio_98, Molmer_93, Kilin_07} of e-h pair, which are the superpositions of deterministic evolutions and random jumps under the action of Lindblad operators. In this case the Hamiltonian \eqref{hamiltonian} should be replaced by the efficient  non-Hermitian Hamiltonian $\hat {H}_{eff} = \hat {H} - i\hbar \sum_p\hat {X}_p^+\hat {X}_p$, where $\hat {X}_p$ is Lindblad operator for $p$-th QD. Suppose that the density of phonon states of the bath has the form of Lorentz line with frequency $\omega_0$ and width $\lambda$. If the bath is weakly coupled with QD then $\hat {X}_p = \sqrt{\lambda/2}\hat{\sigma_p^-}$  (see Refs \cite{Plenio_98, Molmer_93, Kilin_07}). It means that the decay can be taken into account by  substitution $\omega_0\rightarrow\omega_0-i\lambda$ into Eq.\ref{system_diskr_a} and $\omega_0\rightarrow\omega_0+i\lambda$ into Eq.\ref{system_diskr_b}. Let us restrict here consideration to
linear regime of the carrier motion and omit the terms
$O(\Delta\omega)$ in \eqref{system_diskr_a}--\eqref{system_diskr_b}. In the Sec.\ref{sec:local_field} we shall analyze the limitations imposed by omission of nonlinear terms. By this means, omitting the terms $O(\Delta\omega)$, we arrive at the set of difference-differential equations for vectors  $\bm{\Psi}_{p,n}(t) = \left[ \begin{array}{c} A_{p,n}(t) \\ B_{p,n+1}(t) \end{array} \right]$ as follows:

\begin{equation}
\begin{split}
\label{system_diskr2}
\partial_t \bm{\Psi}_{p,n} =\left[-\frac{i\omega_0}{2}\hat{\sigma}_z-\frac{{\lambda}}{2}\hat{I}-ig\sqrt{n+1}\,\hat{\kappa}_p(t)\right] \bm{\Psi}_{p,n}\\
+i\hat{\xi}\left(\bm{\Psi}_{p-1,n}+\bm{\Psi}_{p+1,n}\right),
\end{split}
\end{equation}
where $\hat{I}$ denotes 2D unit operator, $\hat{\kappa}_p(t)=\hat{\sigma}_x \exp[i(\omega t-kpa)\hat{\sigma}_z]$, $\hat{\xi} =[(\xi_1+\xi_2)\hat{I}+(\xi_1-\xi_2)\hat{\sigma}_z]/2$.  Coefficients in \eqref{system_diskr2} are expressed in terms of Pauli matrices
\begin{equation}
\label{Pauli_matrices}
\hat{\sigma}_x=\left(\begin{array}{cr} 0&1 \\ 1&0 \end{array}\right), \hat{\sigma}_y=\left(\begin{array}{cr} 0&-i \\ i&0 \end{array}\right), \hat{\sigma}_z=\left(\begin{array}{cr} 1&0 \\ 0&-1 \end{array}\right),
\end{equation}
acting on vectors  $\bm{\Psi}_{p,n}(t)$.

\subsection{Continuous limit}
\label{sec:cont_lim}
Equations \eqref{system_diskr2}  are the recurrent ordinary differential equations. In some cases it is more convenient to transform them into the system of partial differential equations. To do this we should turn to the continuous limit, making the standard substitutions (see, for example, \cite{Martin_Rothen_02}) $pa \to x$, $\bm{\Psi}_{p,n} \to \bm{\Psi}_n(x)$,  $\bm{\Psi}_{p-1,n}+ \bm{\Psi}_{p+1,n}-2\bm{\Psi}_{p,n} \to a^2\partial^2_x \bm{\Psi}_n$, $\hat{\kappa}_p \to \hat{\kappa}(x)$. Then the system \eqref{system_diskr2} leads to
\begin{equation}
\label{system}
\begin{split}
\partial_t \bm{\Psi}_n = \left[2i\hat{\xi}-\frac{i\omega_0}{2}\hat{\sigma}_z-\frac{{\lambda}}{2}\hat{I}-ig\sqrt{n+1}\,\hat{\kappa}(\omega t-kx)\right]\bm{\Psi}_n \\
+ia^2\hat{\xi}\partial^2_x\bm{\Psi}_n.
\end{split}
\end{equation}

Formula \eqref{system} represents the system of partial differential equations with the variable coefficient $\hat{\kappa}=\hat{\kappa}(\omega t-kx)$. This is the basic system for most of our further calculations.

\section{Traveling Rabi waves}
\label{sec:trav_w}
\subsection{Eigenmodes}
\label{sec:Eigenmodes}

Consider elementary
solution of  system \eqref{system} in the form
of damping traveling waves: $A_n =u_n e^{i(h+k/2)x}e^{-i(\nu+\omega/2)t}e^{-\lambda t/2}$, $B_{n+1}=v_{n+1}e^{i(h-k/2)x}e^{-i(\nu-\omega/2)t}e^{-\lambda t/2}$, where $h$ is a given wave
number and $\nu$ is the eigenfrequency to be found, $u_n$, $v_n$ are the  unknown constant coefficients. Substituting them into equations \eqref{system} and omitting common factors, we obtain
\begin{align}
\label{char_eq_a}
&\left\{\nu-\Delta/2+\vartheta_1(h)\right\}u_n-g\sqrt{n+1}v_{n+1} = 0, \\
\label{char_eq_b}
&g\sqrt{n+1}u_n-\left\{\nu+\Delta/2+\vartheta_2(h)\right\}v_{n+1} = 0,
\end{align}
where $\Delta=\omega_0-\omega$, 
\begin{align}
\label{theta}
\vartheta_{1,2}(h)=\xi_{1,2}[2-a^2(h\pm k/2)^2], 
\end{align}

System \eqref{char_eq_a}--\eqref{char_eq_b} has a nontrivial solution if its determinant is equal to zero, i.e. we obtain quadratic equation which connects $\nu$ and $h$. Solving the  equation with respect to $\nu$, we determine the eigenfrequencies of system as
\begin{equation}
\label{disp_law}
\nu_{_{1,2}}(n,h) = -\frac{1}{2}\left[\vartheta_1(h)+\vartheta_2(h)\mp \Omega_n(h)\right].
\end{equation}
Here
\begin{align}
\label{omega}
\Omega_n(h)=\sqrt{\Delta_{eff}^2+4g^2(n+1)},\\
\label{delta_eff}
\Delta_{eff}(h) = \Delta -\vartheta_1(h)+\vartheta_2(h).
\end{align}
It follows from \eqref{disp_law} that there are two eigenmodes for each value of photon number $n$, namely
\begin{equation}
\label{state_vector2}
\left| {\Psi_{i,n} (t)} \right\rangle =\sum\limits_p \left(A_{i,n}(pa,t) \left|a_p,n \right\rangle +B_{i,n}(pa,t) \left| b_p,n+1 \right\rangle \right) ,
\end{equation}
where $i=1,2$ is the number of the eigenmode, and coefficients $A_{i,n}(x,t)$, $B_{i,n}(x,t)$ are defined by the expressions
\begin{align}
\label{mode1}
\nonumber A_{1,n}(x,t)&=\displaystyle\frac{ C_1 g\sqrt{n+1}e^{i(h+k/2)x}e^{-i(\nu_1+\omega/2-i\lambda/2)t}}{\nu_1-\Delta/2+\vartheta_1}\,, \\
    \rule{0in}{4ex}
B_{1,n+1}(x,t) &= C_1 e^{i(h-k/2)x}e^{-i(\nu_1-\omega/2-i\lambda/2)t}
\raisetag{1pt}
\end{align}
and
\begin{align}
\label{mode2} 
A_{2,n}(x,t) &= C_2 e^{i(h+k/2)x}e^{-i(\nu_2+\omega/2-i\lambda/2)t},\\
    \rule{0in}{6ex}
\nonumber B_{2,n+1}(x,t)&=\displaystyle\frac{C_2 g\sqrt{n+1} e^{i(h-k/2)x}e^{-i(\nu_2-\omega/2-i\lambda/2)t}}{\nu_2+\Delta/2+\vartheta_2},
\end{align}
where $C_{1,2}\equiv C_{1,2}(n,h)$ are normalizing constants, $\vartheta_{1,2}\equiv \vartheta_{1,2}(h)$, the values $\nu_{1,2}\equiv \nu_{1,2}(n,h)$ are expressed by the Eqs. \eqref{theta}, \eqref{disp_law}, respectively. Each mode corresponds to the electron-photon
entangled state, which partial amplitudes oscillate both in time and space.  The expression \eqref{state_vector2} with \eqref{mode1} or \eqref{mode2} describes the \textit{state of radiation-dressed  QD-chain} since the entangled 
state includes field states with different photon numbers ($n$ and $n + 1$).  This states are the generalization of the single atom dressed states \cite{Cohen-Tannoudji_1998} for the case of distributed system. The qualitative distinction of this case is the space-time modulation of dressing parameter: it propagates along the QD-chain according to the traveling wave law $\exp[i(kx-\omega t)]$. Entanglement and dressing are caused by the interaction of light with QD chain and vanish in the limit of $g \rightarrow 0$. In that
case, \eqref{system} describes QD-chain electron-hole pairs in equilibrium and inverse states, respectively.

Spatial oscillations of the partial amplitudes are due to QD-coupling.
They vanish in the limit of $\xi_{1,2} \rightarrow 0$. Generally, both the modes are
excited simultaneously. However, any of them can be excited separately
by a proper choice of initial conditions. Similarly to the other
coherent excitations in condensed matter,  Rabi waves
\eqref{mode1} and \eqref{mode2} introduce a new family of
quasi-particles (we name them \emph{rabitons}). One can apply to them the
standard secondary quantization technique.
Similarly to the effect of the self-induced transparency, the Rabi
wave propagation can be interpreted as the motion of a precessing
pseudo dipole \cite{Shen_84}. However,  the coherence mechanisms in
these two cases are principally different: in the Rabi wave the coherence is
settled by the dispersion relation \eqref{disp_law} while in the
case of the self-induced transparency it has solitonic character.

Note that the eigenmodes (\ref{mode1}) and  (\ref{mode2}) each
comprise traveling waves with \textit{different} wave numbers $h \pm
k/2$. Physically, it means that the Rabi wave propagates in an
effective periodically inhomogeneous medium formed by spatially
oscillating (with period $2\pi/k$) electric field. Therefore, the
diffraction appears in the system.  In the limit $k \rightarrow
0$, the medium turns homogeneous and the diffraction effect vanishes.
Reflections of Rabi waves and their mutual transformations at the field
inhomogeneities become possible. Thus one obtain a unique
possibility to control the processes of the reflection and transmission of
Rabi waves by varying the spatial structure of the light .

\subsection{Dispersion characteristics}
\begin{figure}[htb]
\includegraphics[width=0.48\textwidth,trim=17 15 5 9,clip]{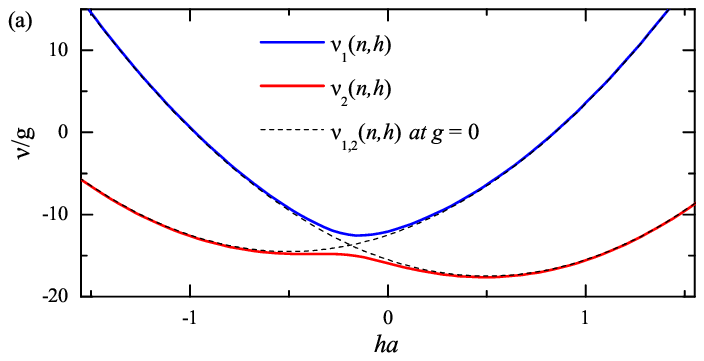}
\includegraphics[width=0.48\textwidth,trim=18.5 14 20 12,clip]{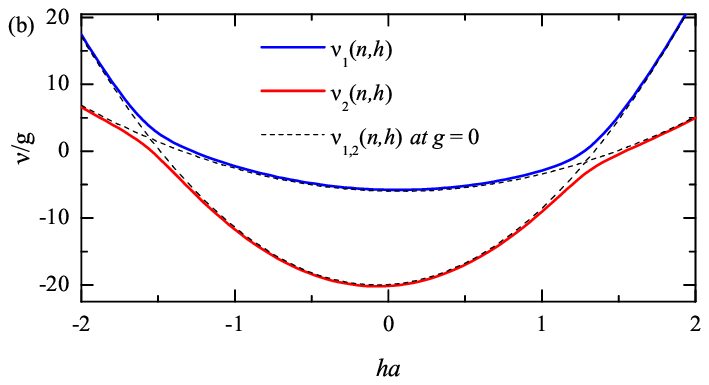}
\includegraphics[width=0.48\textwidth,trim=17 15 9 9,clip]{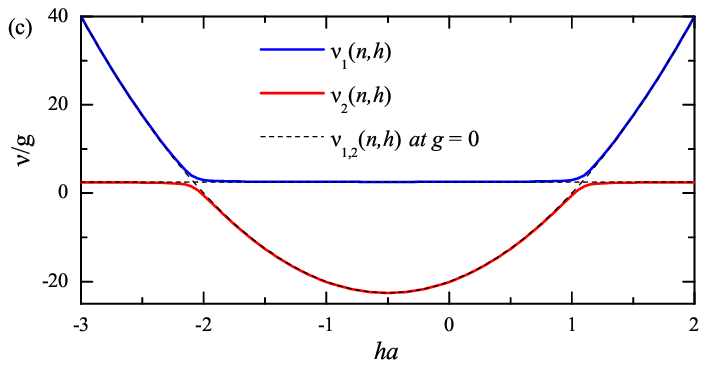}
\caption{\label{fig_disp_char} Dispersion curves for Rabi waves for different parameters values. a) $\xi_1 = \xi_2 =8 g$,  $\Delta = 3g$,  $ka=1$, $n=5$; b) $\xi_1 = 10g$,  $\xi_2 =3 g$,  $\Delta = 0$,  $ka=0.143$, $n=2$; c) $\xi_1 = 10g$,  $\xi_2 =0$,  $\Delta = -5g$,  $ka=1$, $n=5$. }
\end{figure}

Note, that for the waves under consideration \textit{the reciprocity conditions break down}: $\nu_{1,2}(n,h)\neq\nu_{1,2}(n,-h)$.  This is due to the presence of preferential direction, which is determined by the direction of photon mode propagation along the QD-chain (sign of the value $k$). The form of the dispersion characteristics for typical values of parameters is presented on Fig.\ref{fig_disp_char}.
To interpret them one should note that in the limit $g\rightarrow 0$ both of branches have cross points, two in general case and one at $\xi_1=\xi_2$. Modes under examination are characterized by continuous spectrum (the value $h$ varies continuously).

\begin{figure}[htb]
\includegraphics[width=0.48\textwidth,trim=12 15 10 5,clip] {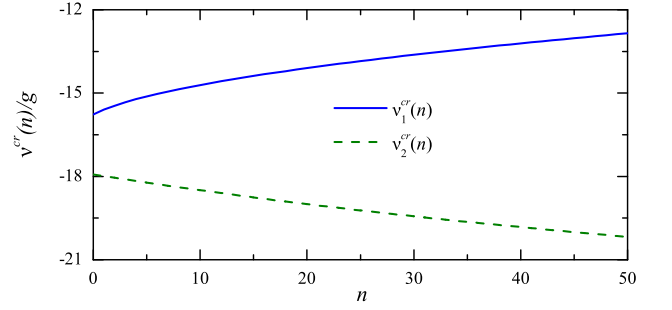}
\caption{\label{critical_fr}  Dependence of critical frequencies of Rabi waves  on the photon number, for the
input parameters as follows: $\xi_1 = 10g$, $\xi_2 =7 g$,  $\Delta = 2(\xi_1-\xi_2)+\xi_2a^2k^2$,  $ka=1$. }
\end{figure}

The values $\nu_{1,2}$ are real for all real $h$. Therefore, the system being investigated is stable \cite{Landau_Lifshitz_1981}. Dispersion equation for \eqref{char_eq_a}-\eqref{char_eq_b} can be solved with respect to wave number $h (\nu)$ for given real-valued frequency $\nu$. In this case for every mode there are such critical values $\nu_{1,2}^{cr}$ that for $\nu<\nu_{1,2}^{cr}$ the values $h_{1,2}(\nu)$ become complex. Physically, it means opacity of the QD-chain for Rabi waves with such frequencies\cite{Landau_Lifshitz_1981}. For $\nu_{2}^{cr}<\nu<\nu_{1}^{cr}$ the QD-chain is opaque only for one of the modes and for $\nu<\nu_{1,2}^{cr}$ it is non-transparent for both of them. One can choose the system parameters in such a way that in some range of $\nu$ the all real solutions $h(\nu)$ are either negative or positive (as an example see Fig.\ref{fig_disp_char}a). This implies one-way opacity of the system (Rabi waves could propagate in one direction only), which is caused by nonreciprocity of the QD-chain. The critical frequencies are determined from the condition $\nu_{1,2}^{cr}(n)=\nu_{1,2}(n, h^{(0)})$, where 
\begin{equation}
\left.\frac{\partial \nu_{1,2}(n, h)}{\partial h}\right|_{h=h^{(0)}_{1,2}} = 0.
\end{equation}
It is significant, that critical values $\nu_{1,2}^{cr}$ depend on photon number $n$ (see Fig.\ref{critical_fr}). This gives an important result, namely: \textit{the conditions of  transparency of the QD-chain are different for Rabi waves with different photon numbers}. Note, that $\nu_{1}^{cr}$ increases and $\nu_{2}^{cr}$ decreases with increasing of $n$.

 Usually both Rabi waves demonstrate the normal dispersion and propagate in the \textit{same direction}. But for a special choice of the parameters the \textit{abnormal dispersion region} for one of the modes appears. For this mode phase and group velocities are oppositely directed  (see Fig.\ref{fig_disp_char}a). The dispersion characteristics have special behavior for absolutely nontransparent ground-state barrier: $\xi_2=0$ (Fig.\ref{fig_disp_char}c). In this case the group velocity $v_{gr}=\partial \nu/\partial h$ of one of the waves is very small for all $h$, except for the narrow vicinities of the cross-points. Due to this fact the modes exchange their places when passing the cross points (e.g. on Fig.\ref{fig_disp_char}c $v_{gr} \approx 0$ for $\nu_1$-mode between cross points $-2 \apprle ha \apprle 1$ and for $\nu_2$-mode at all other values of $h$). Thus, in  current case for a wide range of $h$ one of the modes practically does not transfer the energy along the QD-chain.

\section{Rabi wavepackets}
\label{sec:wave_pack}
\subsection{Electron-photon dynamics}

To find  the general solution of the system \eqref{system} let us 
introduce the new variable $\bm{\Phi}_n(x,t)=e^{i(\omega_0 t - kx)\hat{\sigma}_z/2}e^{\lambda t/2}\bm{\Psi}_n(x,t)$.
It leads to the system of partial differential equations with coefficients independent of QD-position $x$:

\begin{equation}
\begin{split}
\label{system2}
\displaystyle \partial_t\bm{\Phi}_n-i \left[\left(2-a^2k^2/4\right)\hat{\xi}-g\sqrt{n+1}\hat{\chi}(t)\right]&\bm{\Phi}_n\\
+a^2k\hat{\sigma}_z\hat{\xi}\partial_x\bm{\Phi}_n-ia^2&\hat{\xi}\partial_x^2\bm{\Phi}_n=0,
\end{split}
\end{equation}
where $\hat{\chi}(t)=\hat{\sigma}_x \exp(-i\hat{\sigma}_z\Delta t)$. This system can be solved by using the Fourier transform with respect to $x$:
\begin{align}
\label{fourier_tr}
\bm{\Phi}_n(x,t)=\int\limits_{-\infty }^\infty {\widetilde{\bm{\Phi}}_n(h,t)e^{ihx}} dh. 
\end{align}
Substituting Eq.\eqref{fourier_tr} into Eq.\eqref{system2}, we obtain the system of ordinary differential equations

\begin{align}
\label{system3}
\frac{d}{dt}\widetilde{\bm{\Phi}}_n=i\left(\hat{\vartheta} -g\sqrt{n+1}\hat{\chi}(t)\right)\widetilde{\bm{\Phi}}_n,
\end{align}
where $\hat{\vartheta}=\hat{\vartheta}(h)=[(\vartheta_1(h)+\vartheta_2(h))\hat{I}+(\vartheta_1(h)-\vartheta_2(h))\hat{\sigma}_z]/2$. The system \eqref{system3} can be easily integrated, the solution is given by
\begin{align}
\label{sol}
\widetilde{\bm{\Phi}}_n(h,t)=\hat{\rho}_n(h,t)\widetilde{\bm{\Phi}}_n(h,0),
\end{align}
where
\begin{equation}
\label{rho}
\hat{\rho}_n(h,t)=e^{i[\vartheta_1(h)+\vartheta_2(h)]t}e^{i\hat{\sigma}_z \Delta t} e^{i[\bm{m}\hat{\bm{\sigma}}\Omega_n(h)]t},
\end{equation}
$\hat{\bm{\sigma}} = (\hat{\sigma}_x, \hat{\sigma}_y, \hat{\sigma}_z)$, $\bm{m}$ is the unit vector of the form $\left(-2g\sqrt{n+1}/\Omega_n, 0, \Delta_{eff}/\Omega_n \right)$. The expression \eqref{rho} can be written in matrix form as follows:
\begin{equation}
\label{rho_2}
\hat{\rho}_n(h,t)=\left[\begin{array}{cc} {\varphi_n^-(h,t)e^{i\delta_+(h)t}}&{\psi_n(h,t)e^{i\delta_+(h)t}} \\ {\psi_n(h,t)e^{i\delta_-(h)t}}&{\varphi_n^+(h,t)e^{i\delta_-(h)t}} \end{array}\right],
\end{equation}

\begin{align}
\label{phi}
\varphi_n ^\pm (h,t)=\cos \frac{\Omega_n(h) t}{2} \pm i\frac{\Delta_{eff}(h)}{\Omega_n(h)}\sin \frac{\Omega_n(h) t}{2},\\
\label{psi}
\psi_n (h,t)=-i\frac{2g\sqrt{n+1}}{\Omega_n(h)} \sin \frac{\Omega_n(h) t}{2},
\\
\label{delta}
\delta_{\pm}(h) =\frac{1}{2}\left[\vartheta_1(h)+\vartheta_2(h)\pm\Delta\right], 
\end{align}
$\Omega_n (h)$ and $\Delta_{eff}(h)$ are determined by \eqref{omega} and \eqref{delta_eff}, correspondingly, and $\widetilde{\bm{\Phi}}(h,0)$ can be obtained from initial conditions by the inverse Fourier transform:
\begin{align}
\label{inv_f_transform_2}
\widetilde{\bm{\Phi}}_n(h,0)=\frac{1}{2\pi }\int\limits_{-\infty }^\infty {e^{-i(h \hat{I}+\hat{\sigma}_zk/2)x}\bm{\Psi}_n(x,0)} dx.
\end{align}

Thus, for the  vector of probability amplitudes $\bm{\Psi}_n(x,t)$ we have the following expression:
\begin{equation}
\begin{split}
\raisetag{50pt}
\label{sol2}
\bm{\Psi}_n(x,t)=&e^{i(kx-\omega t)\hat{\sigma}_z/2}e^{-\lambda t/2}\\
\times&\int\limits_{-\infty }^\infty {\hat{\rho}_n(h,t)\widetilde{\bm{\Phi}}_n(h,0) e^{i\{hx+[\vartheta_1(h)+\vartheta_2(h)]t/2\}}} dh.
\end{split}
\end{equation}

Expression \eqref{sol2} is the central result of this section. It allows one to characterize such observables  as inversion, as well as the exciton-exciton and exciton-photon correlators for different initial states of light and QDs.

\subsection{Dispersionless approximation}
\label{sec:dispersionless_app}

In this section we will analyze the propagation of Rabi waves with spatially localized initial state. The QD-chain  issupposed to be initially either in the ground state ($A_n(x,0)=0$ for all $n$), or in the excited state   ($B_n(x,0)=0$ for all $n$), or in the  mixed state (arbitrary superposition of ground and excited states). 

We describe the spatial-temporal dynamics of Rabi oscillations  by the spatial density of the  inversion  (the inversion per single QD)
\begin{equation}
\label{inv_density}
w(x,t)=a\sum \limits_n[|A_n(x,t)|^2-|B_{n+1}(x,t)|^2].
\end{equation}
Space-averaged temporal evolution of Rabi oscillations is characterized by the integral inversion

\begin{equation}
\label{int_inversion}
\widetilde{w}(t)=\int \limits_{-\infty}^{\infty}w(x,t)dx.
\end{equation}

Let us approximate the  initial spatial distributions  by the Gaussian beams $C_{A,B}\exp[-(x-d_{A,B})^2/2\sigma_{A,B}^2]$, where $C_{A,B}$ are normalization constants, $d_{A,B}, \sigma_{A,B}$ are the positions of the beams and their widths, respectively (the indexes $A, B$ refer to excited and ground state of the electron in the QD-chain, respectively). 

The Rabi wavepacket evolution can be qualitatively illustrated with the help of the dispersionless approximation for Eqs.\eqref{sol2}. According to  \eqref{inv_f_transform_2} we obtain for $\widetilde{\bm{\Phi}}_n(h,0)=\left[\begin{array}{c} {a_n(h)}\\{b_{n+1}(h)} \end{array} \right]$:
\begin{align}
\label{init_f_a}
a_n (h)\sim c(n) e^{-(h+k/2)^2\sigma_A^2/2}e^{-i d_A(h+k/2)},\\
  \rule{0in}{6ex}
  \label{init_f_b}
b_{n+1}(h)\sim c(n+1) e^{-(h-k/2)^2\sigma_B^2/2}e^{-i d_B(h-k/2)},
\end{align}
where $c(n)$ is an arbitrary photonic distribution. One can see from \eqref{init_f_a}--\eqref{init_f_b} that the spatial spectrum of $\bm{\Psi}_n(x,0)$ is given by two Gaussian peaks with peak
widths $1/\sigma_{A,B}$ and positions of the peak centers $h_{1,2}^{(0)}=\pm k/2$, respectively. If $\sigma_{A,B}$ are sufficiently large, the largest contribution to the integration in \eqref{sol2} comes from the narrow vicinities of points  $h_{1,2}^{(0)}$. 
 Assuming $\vartheta_i(h)\cong\vartheta_i(h^{(0)}_{1,2})+\left.(\partial\vartheta_i/\partial h)\right|_{h=h^{(0)}_{1,2}}(h-h^{(0)}_{1,2})$ and $\hat{\rho}_n(h,t)\cong \hat{\rho}_n(h^{(0)}_{1,2},t)$ (the dispersion effects neglecting) and doing standard calculations, we obtain from Eq.\eqref{sol2}:

\begin{widetext}
\begin{align}
\label{appr_sol_a}
\nonumber A_n(x,t)&=e^{i(kx-\omega t)/2}e^{-\lambda t} \left\{\left[A_{n}(x+v_2^+t,0)\zeta_2^+e^{i\theta_2^+t}+A_{n}(x+v_2^-t,0)\zeta_2^-e^{i\theta_2^-t}\right]e^{-ikx/2}\right.\\
 &+\left.\left[B_{n+1}(x+v_1^+t,0)\eta_1^-e^{i\theta_1^+t}+B_{n+1}(x+v_1^-t,0)\eta_1^+e^{i\theta_1^-t}\right]e^{ikx/2}\right\},\\
 \label{appr_sol_b}
\nonumber B_{n+1}(x,t)&=e^{-i(kx-\omega t)/2}e^{-\lambda t} \left\{\left[A_{n}(x+v_2^+t,0)\eta_2^-e^{i\theta_2^+t}+A_{n}(x+v_2^-t,0)\eta_2^+e^{i\theta_2^-t}\right]e^{-ikx/2}\right.\\
&+\left.\left[B_{n+1}(x+v_1^+t,0)\zeta_1^-e^{i\theta_1^+t}+B_{n+1}(x+v_1^-t,0)\zeta_1^+e^{i\theta_1^-t}\right]e^{ikx/2}\right\},
\end{align}
\end{widetext}
where the velocities $v^{\pm}_{1,2}$ are defined as
\begin{equation}
\label{group_v_a}
v^{\pm}_1=v^{\pm}(h_1^{(0)})=-\xi_1a^2k\frac{\Omega_n(k/2) \mp \Delta_{eff}(k/2)}{\Omega_n(k/2)},
\end{equation}
\begin{equation}
 \label{group_v_b}
 v^{\pm}_2=v^{\pm}(h_2^{(0)})=\xi_2a^2k\frac{\Omega_n(-k/2) \pm \Delta_{eff}(-k/2)}{\Omega_n(-k/2)},
\end{equation}
and $\theta_{1,2}^{\pm}$ and $\zeta_{1,2}^{\pm}$, $\eta_{1,2}^{\pm}$ are introduced to denote the frequency shifts and amplitude factors, respectively:
\begin{align}
\label{theta_1_2}
\theta_{1,2}^{\pm}&=\frac{1}{2}[\vartheta_1(h_{1,2}^{(0)})+\vartheta_1(h_{1,2}^{(0)})+\Omega_n(h_{1,2}^{(0)})]-v^{\pm}(h_{1,2}^{(0)}) , \
\\
\label{zeta}
\zeta_{1,2}^{\pm}&=\frac{\Omega_n(h_{1,2}^{(0)}) \pm \Delta_{eff}(h_{1,2}^{(0)})}{2\Omega_n(h_{1,2}^{(0)})}, \
\\
\label{eta}
\eta_{1,2}^{\pm}&=\pm \frac{g\sqrt{n+1}}{\Omega_n(h_{1,2}^{(0)})}. \ \
\end{align}
 As is apparent from \eqref{appr_sol_a}--\eqref{appr_sol_b}, generally, any probability amplitude in the Rabi-wave packet is the \textit{superposition of four subpackets}. Two of them (first and second terms in expressions \eqref{appr_sol_a}--\eqref{appr_sol_b}) correspond to the excited initial state and two another (third and fourth terms in \eqref{appr_sol_a}--\eqref{appr_sol_b}) correspond to the ground initial state. The partial subpackets in Eqs.~\eqref{appr_sol_a}--\eqref{appr_sol_b} are characterized by the different frequency shifts $\theta_{1,2}^{\pm}$ and different velocities of motion  $v^{\pm}_{1,2}$, given by \eqref{group_v_a}, \eqref{group_v_b}. This velocities coincide with the group velocities of traveling Rabi waves \eqref{mode1}--\eqref{mode2} when $h=\pm k/2$: one can elementary verify, that  $v^{\pm}_1=\left.\partial\nu_{1,2}/\partial h\right|_{h=k/2}=v_{1,2}^{gr}(k/2)$, $v^{\pm}_2=\left.\partial\nu_{1,2}/\partial h\right|_{h=-k/2}=v_{1,2}^{gr}(-k/2)$. 
 It is essential that velocities  $v^{\pm}_{1,2}$ as well as the frequency shifts $\theta_{1,2}^{\pm}$ depend  on photon number $n$. It means that the spatial propagation of wavepacket  is accompanied by the change of quantum light statistics (for example, in initially coherent light incoherent component appears). 
 
 Under the some specific conditions the number of subpackets could decrease. Two mechanisms of such decrease are possible:  the first one is the tending to zero the subpacket amplitude and the second one is the confluence of the subpackets velocities. 
 
 As is seen from \eqref{group_v_a}--\eqref{group_v_b}, if 
\begin{equation}
\label{synchr_cond}
    \Delta_{eff}(h_{1,2}^{(0)})=0,
\end{equation}
for one pair of the subpackets the velocity synchronism condition  $v^{+}_{1,2}=v^{-}_{1,2}$  is fulfilled. Notice that Eq.~\eqref{synchr_cond} is the analogue of the ordinary synchronism condition in single two-level system  $\Delta=0$.  The velocity synchronism condition can be fulfilled by fitting the value of the detuning  $\Delta$. If this is the case, there are three subpackets instead of four them due to the confluence mechanism acting. Note, that $\Delta_{eff}(k/2)$ and $\Delta_{eff}(-k/2)$ can not be equal to zero together at $k\neq0$. 

If the system is initially prepared in the stationary state, and the synchronism condition is fulfilled in the corresponding point (for example  $B_{n+1}(x,0)=0$ for any $n$ and $\Delta_{eff}(-k/2)=0$), only one subpacket is preserved. In this case the expression for the inversion density becomes rather simple: 
\begin{equation}
\label{inv_density_2}
w(x,t)=a\sum \limits_n A_n^2(x+\xi_2a^2kt,0)[1-2\sin^2(g\sqrt{n+1}t)].
\end{equation}
Expression \eqref{inv_density_2} sums up contributions of different photonic states (terms with different $n$). Each contribution is the product of two multipliers describing different optical processes. The second multiplier describes Rabi oscillations  with the frequencies $\nu_n = 2g\sqrt{n+1}$, occurring in the QD-chain, while the first one shows that the region of Rabi oscillations  moves therewith in space with velocity $v=\xi_2a^2k$. One should note, that in the case of the exact synchronism $v$ does not depend on $n$. That is why in this partial case the quantum statistics of light is not distorted with propagation of the wavepacket. In particular, the initial coherency of quantum light persists in time. It is immediately follows from Eq.\eqref{int_inversion} and normalization condition that the integral inversion $\widetilde{w}(t)$ corresponding to the inversion density \eqref{inv_density_2} oscillates in time similarly to the Rabi-oscillations in the single two-level system  (see Eq.(6.2.21) in \cite{Scully} at $\Delta = 0$).

Generally, each pair of the subpackets propagates due to the tunneling through its own potential barrier. The velocities of the subpackets motion are proportional to the barrier transparency ($v_{1,2}^{\pm}\sim\xi_{1,2}$). If one of the barriers becomes absolutely opaque, the corresponding pair of subpackets does not move: $v_{1,2}^{\pm}\longrightarrow0$ at $\xi_{1,2}\longrightarrow0$. The light propagation  \textit{along the QD-chain} is another necessary condition of Rabi subpackets  motion: $v_{1,2}^{\pm}\longrightarrow0$ at $k\longrightarrow0$.

For more detail analysis of Rabi wavepackets dynamics (in particular for taking into account the diffraction spreading) the integrals \eqref{appr_sol_a}--\eqref{appr_sol_b} need to be calculated numerically. The results of calculations is analysed in the following section.

\subsection{Coherent state: collapses and revivals picture}
\label{sec:coh_state}

\begin{figure}[ht]
\includegraphics[width=0.48\textwidth,trim=5 5 5 5,clip]{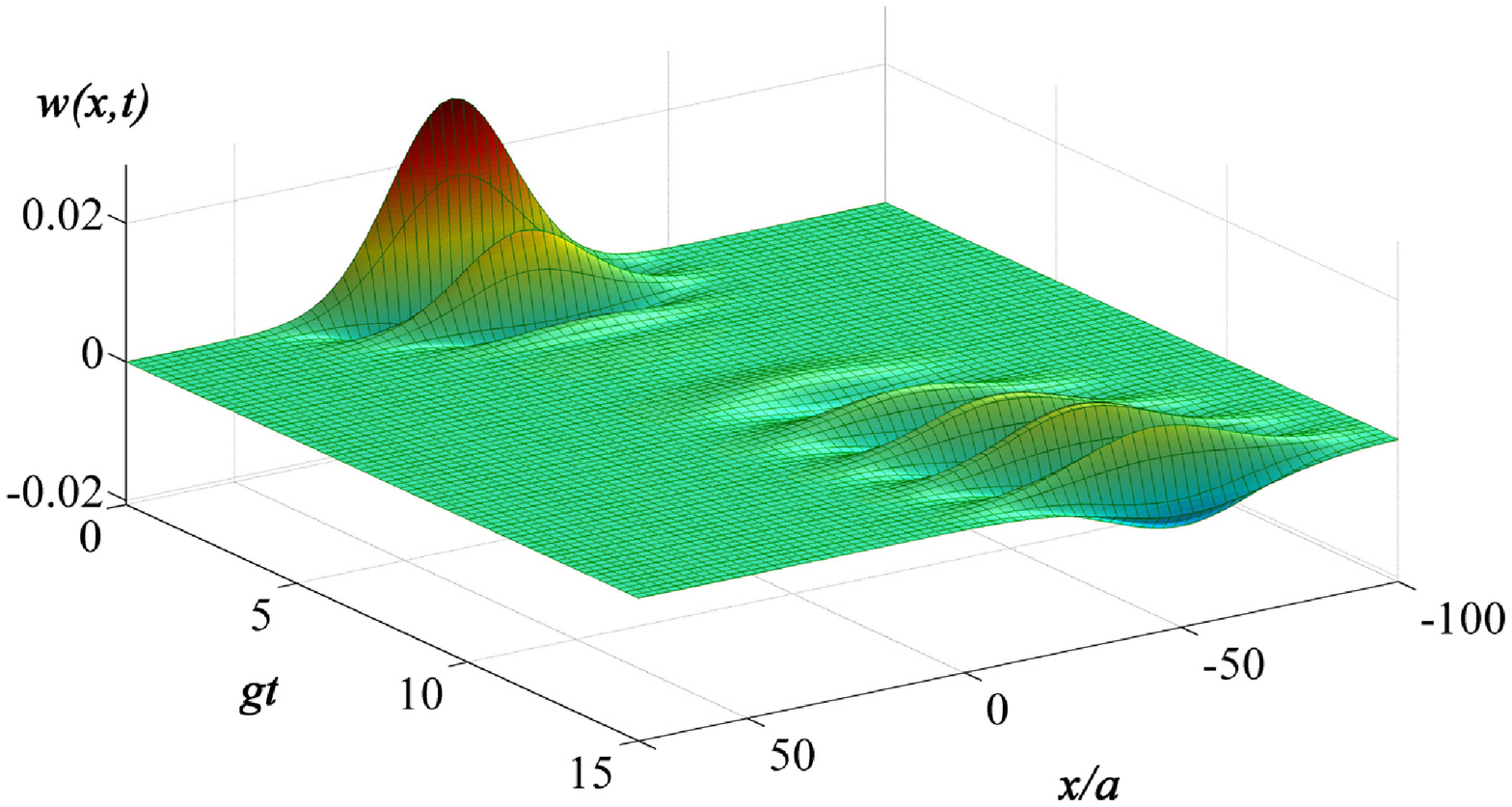}
\includegraphics[width=0.48\textwidth,trim=15 17 15 10,clip]{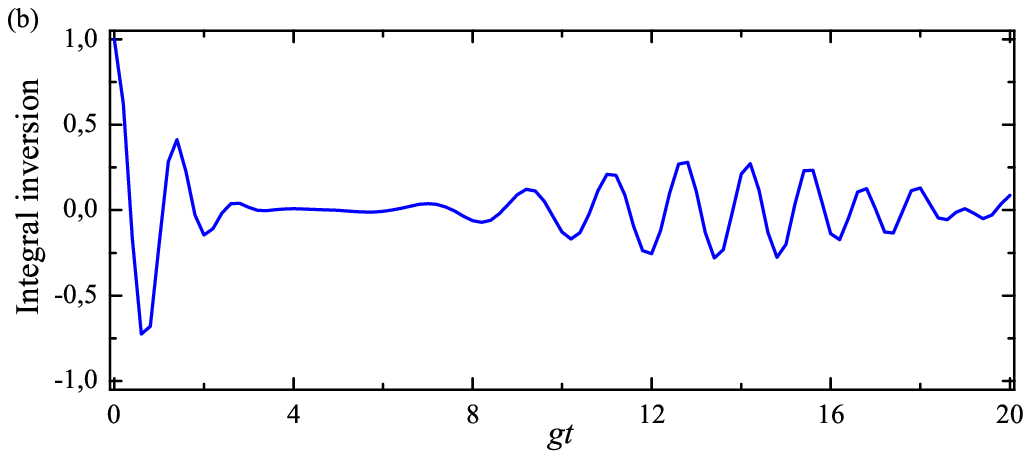}
\caption{\label{fig1} Space-time distribution of the inversion
density (a) and temporal dependence of the integral inversion (b) in the QD chain for a coherent initial state of light ($\left\langle n\right\rangle=4$). The initial state of QD-chain is a single Gaussian wavepacket.
$A_n(x,0)=c(n)\exp(-x^2/2\sigma^2)/\sqrt[4]{\pi\sigma^2}$, $B_{n+1}(x,0)=0$,
$\xi_1=10g$, $\xi_2=7g$, $\Delta = 2(\xi_1-\xi_2)+\xi_2a^2k^2$,
$ka=0.5$, $\sigma = 20 a$,  $\lambda = 0.05g$.}
\end{figure}

Consider the case of an excited initial state with space distribution in form of single Gaussian beam: $A_n(x,0)=c(n)\exp(-x^2/2\sigma^2)/\sqrt[4]{\pi\sigma^2}$, $B_{n+1}(x,0)=0$.  Assume that initially light is prepared in coherent state, so photon distribution is given by the Poisson law: $c(n) = \left\langle n\right\rangle^{n/2}e^{-\left\langle n\right\rangle/2}/\sqrt{n!}$, where $\left\langle n\right\rangle$ defines the average photon number. The spatial-temporal dynamics of inversion density for this case
 is depicted on Fig.\ref{fig1}a. As is seen from it, original Gaussian packet
oscillates in time and moves along the chain. Oscillations  collapse to zero quickly,
but after a while they revive \textit{in another area of space}. The phenomenon of Rabi
oscillations collapses and revivals, caused by discreteness of photon distribution, is well
studied \cite{Scully}, but  the spacing the collapse and subsequent revival is qualitatively new effect.

\begin{figure}[ht]
\includegraphics[width=0.48\textwidth]{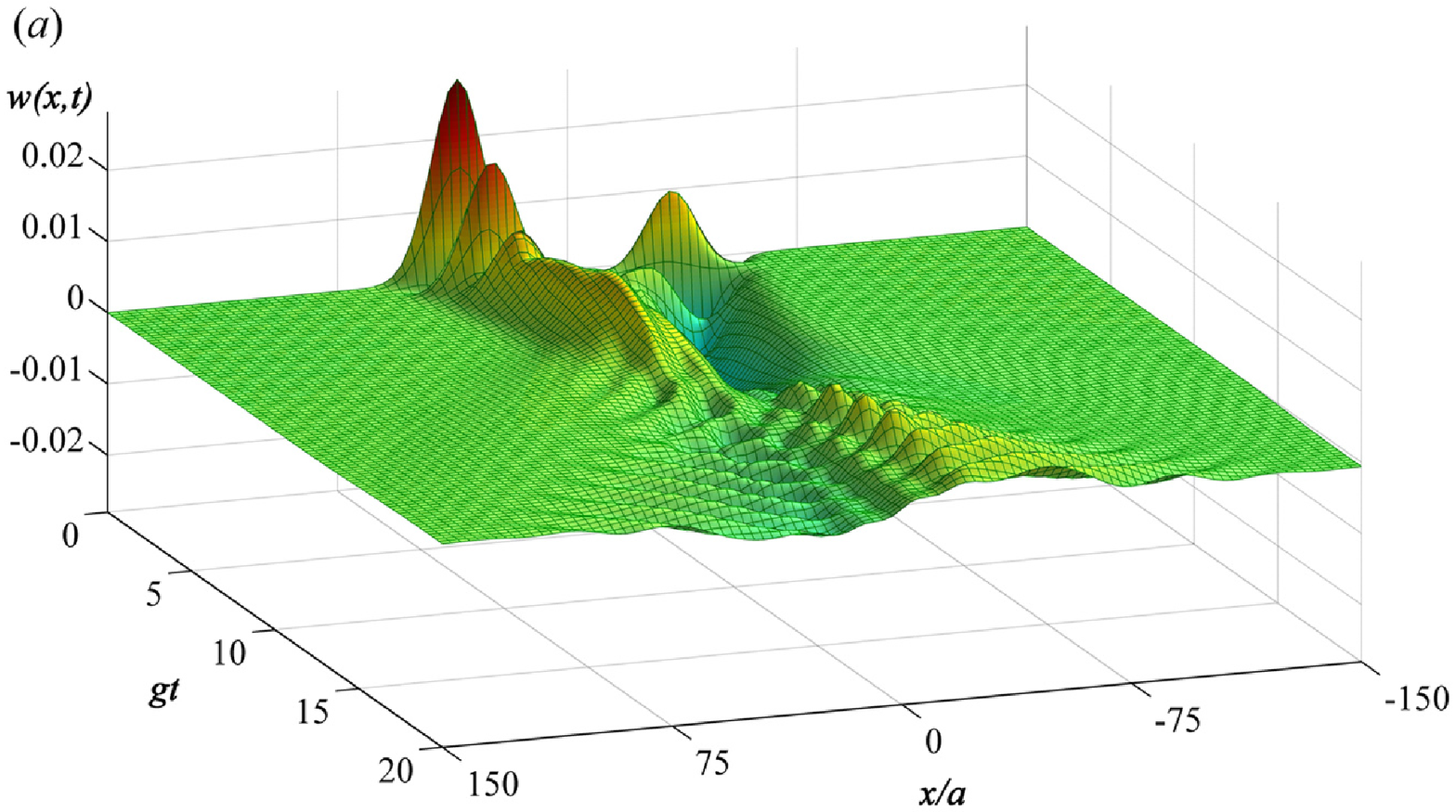}
\includegraphics[width=0.48\textwidth]{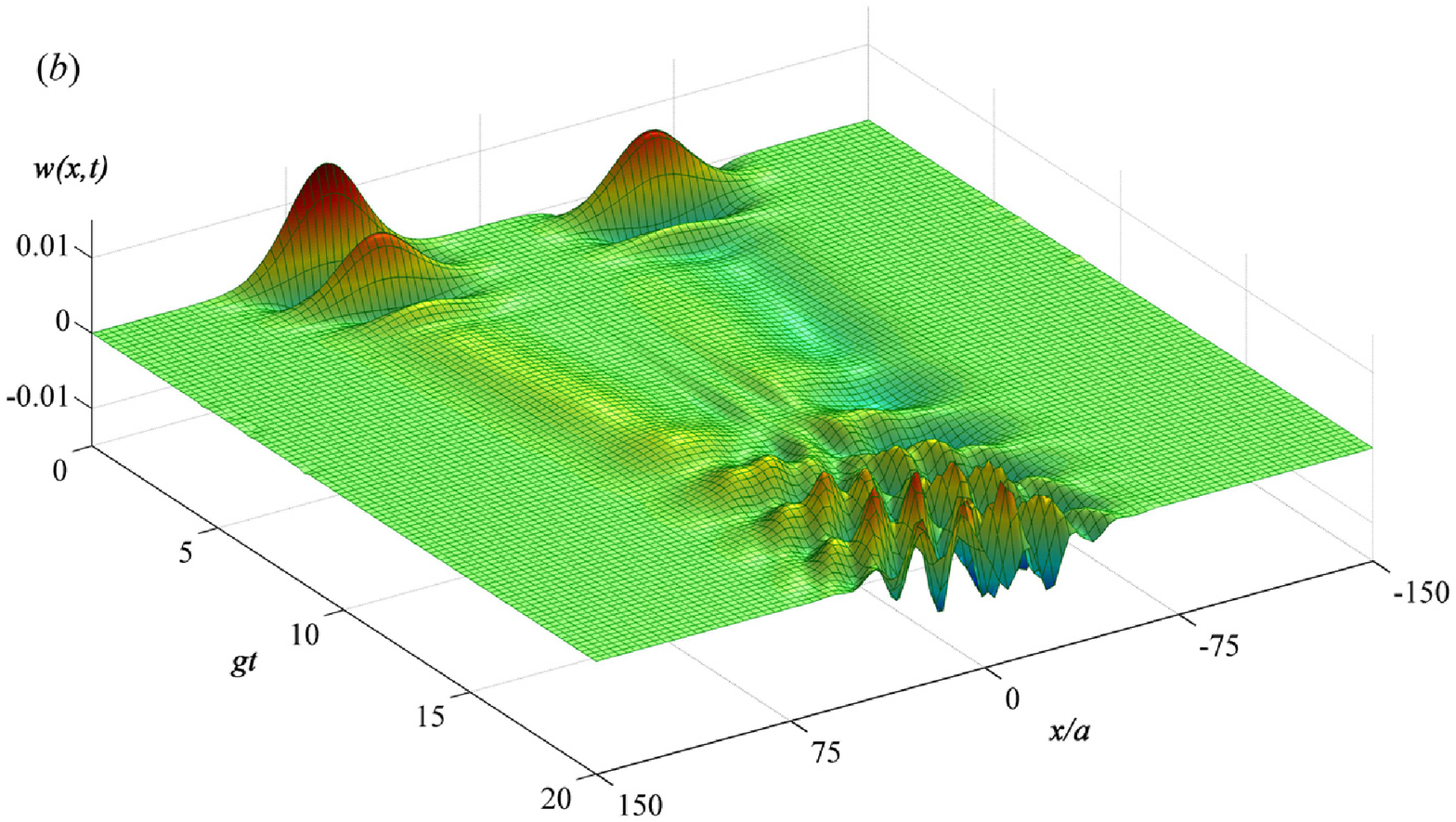}
\caption{\label{double} Space-time distribution of the inversion in
the QD chain for the case of two counterpropagating wavepackets. $A_n(x,0) =
c(n)\exp[-(x-3\sigma)^2/2\sigma^2]/\sqrt[4]{4\pi\sigma^2}$, $B_{n+1}(x,0) =
c(n+1)\exp[-(x+3\sigma)^2/2\sigma^2]/\sqrt[4]{4\pi\sigma^2}$. The light is initially in coherent state ($\left\langle n\right\rangle=5$), $\Delta =
0$ , $\xi_1=\xi_2=10g$, $ka=0.33$. a) $\sigma = 10 a$, $\lambda = 0$; b) $\sigma = 20 a$, $\lambda = 0.05g$.}
\end{figure}

Although  a variation of the inversion density, depicted in Fig.\ref{fig1}a, in arbitrary point of the space occupied by the Rabi
wavepacket is not too large, the integral inversion, presented in Fig.\ref{fig1}b oscillates between -1 and 1, thus indicating presence of strong light-QD coupling.
\begin{figure}[bp]
\includegraphics[width=0.48\textwidth,trim=15 17 15 10,clip]{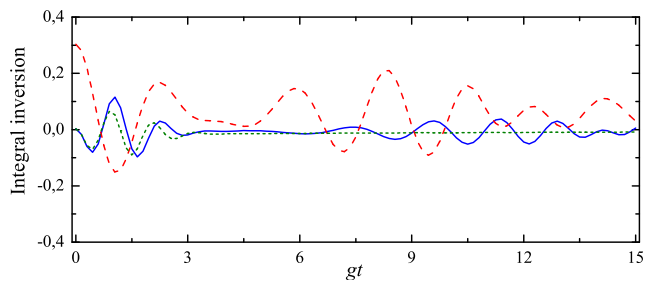}
\caption{\label{int_inv2} Temporal dependence of the  integral inversion for a coherent initial state of light 
at the input parameters as follows: $\left\langle n\right\rangle=5$, $\sigma = 20 a$, $\lambda = 0.05g$  (solid line); $\left\langle n\right\rangle=5$, $\sigma = 10 a$, $\lambda = 0$(dotted line) $\left\langle n\right\rangle=0.5$, $\sigma = 20 a$, $\lambda = 0$(dashed line). In all cases, $\xi_1=\xi_2=10g$, $\Delta =
0$,  $ka=0.33$.  }
\end{figure}
\begin{figure}[ht]
\includegraphics[width=0.48\textwidth,trim=35 15 25 25,clip]{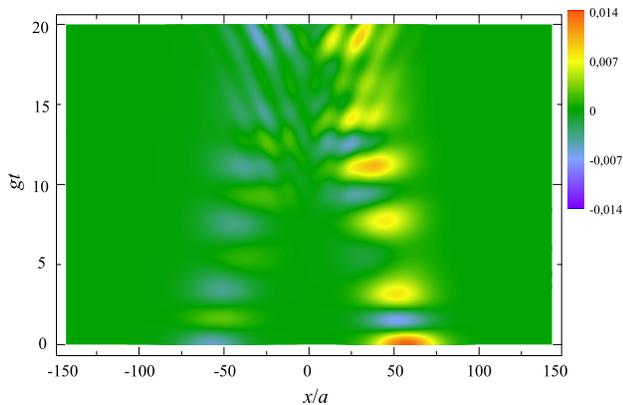}
\caption{\label{double_3} The asymmetry  of the inversion space-time distribution  for two counterpropagating wavepackets in case of the coherent state of light with
 small average number of the photons $\left\langle n\right\rangle=0.5$. $A_n(x,0) =
c(n)\exp[-(x-3\sigma)^2/2\sigma^2]/\sqrt[4]{4\pi\sigma^2}$, $B_{n+1}(x,0) =
c(n+1)\exp[-(x+3\sigma)^2/2\sigma^2]/\sqrt[4]{4\pi\sigma^2}$, $\xi_1=\xi_2=10g$, $\Delta =
0$,  $ka=0.33$, $\sigma = 20 a$, $\lambda = 0$.}
\end{figure}

Let us consider now the case of a mixed initial state (both $A_n(x,0)$ and $B_{n+1}(x,0)$ are nonzero). As it has been noted above, the synchronism condition can not be fulfilled for both pair of subpackets (except the case when one of the pairs is motionless, see below). As a consequence, we have four subpackets (or three ones if $\Delta_{eff}=0$ for one of the pairs). 
If $\xi_1=\xi_2=\xi$ (potential barriers with equal transparency) and $\Delta=0$, then both packet pairs move with pairwise equal velocities in opposite directions ($v_1^{\pm}=-v_2^{\pm}$).  If the initial spatial distributions of probability amplitudes for ground and excited states are spaced, the subpackets will collide (see Fig.\ref{double}). Initial Gaussian profiles of packets deform with time due to the difference in subpacket velocities.

One should note that for the case of quantum external field in coherent state QD-chain cannot be saturated, i.e. the integral inversion could not be a constant (see Fig. \ref{int_inv2}). This is due to the fact that $A_n(x,t)$ and $B_{n+1}(x,t)$ are asymmetrical with respect to $n$ and as a result $\sum_n\int_{-\infty}^{\infty}|A_n(x,t)|^2dx-\sum_n\int_{-\infty}^{\infty}|B_{n+1}(x,t)|^2dx$ is not a constant for any form of space dependence of $A_n(x,t)$, $B_{n+1}(x,t)$. This asymmetry exists due to the presence of vacuum term in the sum and increases with decreasing of the average  photon number $\left\langle n\right\rangle$ (compare Fig.\ref{double} and Fig.\ref{double_3}).

\begin{figure}[htbp]
\includegraphics[width=0.48\textwidth]{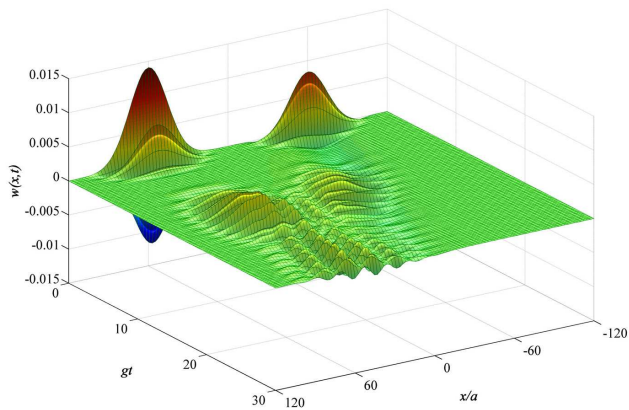}
\caption{\label{double2} Space-time distribution of the inversion in
the QD chain for the case of the fully opaque ground-state barrier. $A_n(x,0) =
c(n)\exp[-(x-3\sigma)^2/2\sigma^2]/\sqrt[4]{4\pi\sigma^2}$, $B_{n+1}(x,0) =
c(n+1)\exp[-(x+3\sigma)^2/2\sigma^2]/\sqrt[4]{4\pi\sigma^2}$, $\xi_1=10g$, $\xi_2=0$, $\Delta =
2(\xi_1-\xi_2)-\xi_1a^2k^2$,  $ka=0.5$, $\sigma = 20 a$, $\lambda = 0.05g$. The light is initially in the coherent state with  $\left\langle n\right\rangle=5$} 
\end{figure}

In other limiting case, for $\xi_2=0$ (fully opaque ground-state barrier), the velocities  $v_2^{\pm}=0$ and one pair of subpackets do not moves along the chain. If, moreover, $\Delta_{eff}(k/2)=0$, then $v_1^{+}=v_1^{-}$, and the synchronism condition is fulfilled for both packet pairs (see Fig.\ref{double2}).

\subsection{Fock qubit state and vacuum Rabi waves}
\label{sec:fock_vacuum}

\begin{figure}[ht]
\includegraphics[width=0.48\textwidth,trim=35 15 28 25,clip]{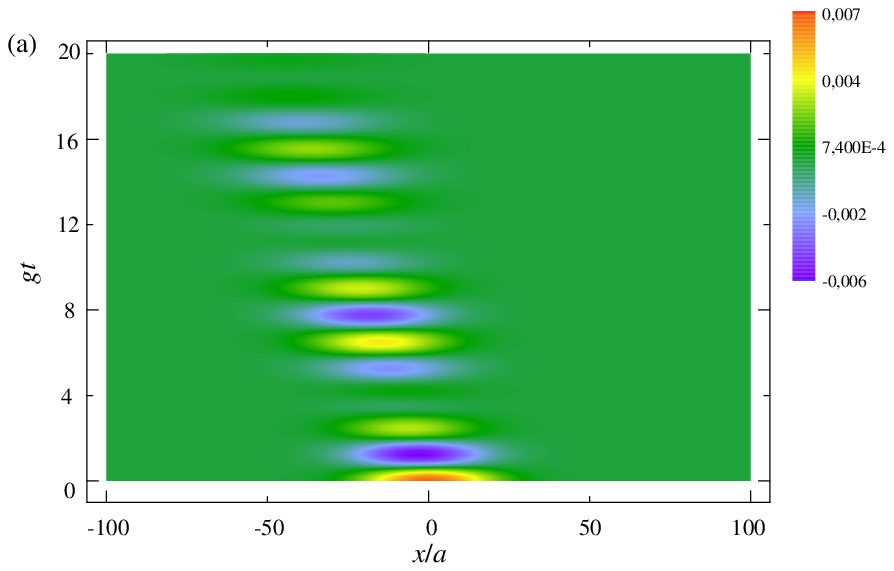}
\includegraphics[width=0.48\textwidth,trim=15 17 15 10,clip]{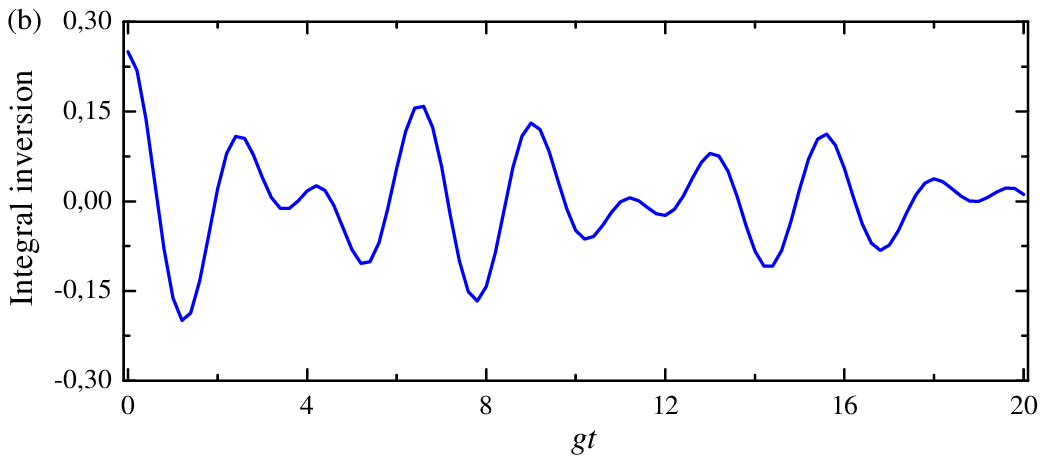}
\caption{\label{fig_fock_qubit} Space-time distribution of the inversion (a) and temporal dependence of the integral inversion (b) in
the QD chain for the field in the Fock qubit initial state. $\left|\psi_{f}(0)\right\rangle=1/\sqrt{2}\left|0\right\rangle+1/\sqrt{2}\left|1\right\rangle$, $A_0(x,0)=A_1(x,0)=(1/\sqrt[4]{\pi\sigma^2})$ $ \exp(-x^2/2\sigma^2)$, $A_n(x,0)=0$ for $n\geq2$, $B_n(x,0)=0$ for all $n$, $\xi_1=10g$, $\xi_2=7g$,
$\Delta = 2(\xi_1-\xi_2)+\xi_2a^2k^2$, $\sigma = 20 a$, $ka=0.33$, $\lambda = 0.05g$.}
\end{figure}

Let us now consider the interaction of the QD-chain with another important state of initial electromagnetic field, namely, the Fock qubit state, i.e. a superposition of two Fock states: $\left|\psi_{f}(0)\right\rangle=C_1\left|N\right\rangle+C_2\left|N+1\right\rangle$, where $N$  is an arbitrary fixed number, $C_{1,2}$ satisfy the normalization conditions. As in the previous case, for initial spatial distributions the Gaussian beams will be used. The spatial-temporal dynamics of the inversion density for the initial  state of field $\left|\psi_{f}(0)\right\rangle=1/\sqrt{2}\left|0\right\rangle+1/\sqrt{2}\left|1\right\rangle$ is shown in Fig. \ref{fig_fock_qubit}a. It can be seen from the figure that Rabi wavepacket does not collapse and revive, but oscillates in a complicated manner (see also integral inversion dynamics, Fig.\ref{fig_fock_qubit}).

In the end of this section we consider the vacuum initial state of external electromagnetic field. It is well known that for this case the vacuum harmonic Rabi oscillations occuring due to spontaneous emission can take place \cite{Scully}. The corresponding Rabi-wave packet in the initially excited QD-chain is illustrated at Fig.\ref{fig_fock_qubit}. This excitation couples single mode electromagnetic vacuum state with one-photon state only and can be imagined as a wave beam which is described by the monochromatic Rabi-frequency spectrum and continuous spatial spectrum at the same time.  However, it should be noted that one need to consider an interaction of the QD-chain with a multi-mode field  for full treatment of such effects \cite{ref01}. This problem is a subject for future considerations.

\begin{figure}[ht]
\includegraphics[width=0.48\textwidth,trim=35 15 28 25,clip]{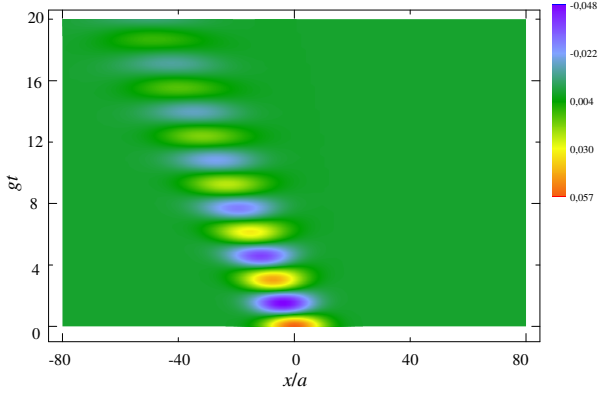}
\caption{\label{fig_vacuum_RW} Space-time distribution of the inversion in
the QD chain for the vacuum initial state of the field.  $A_0(x,0)
=(1/\sqrt[4]{\pi\sigma^2})$ $ \exp(-x^2/2\sigma^2)$ , $A_n(x,0)=0$ for $n\geq1$, $B_n(x,0)=0$ for all $n$, $\xi_1=6g$, $\xi_2=5g$,
$\Delta = 2(\xi_1-\xi_2)+\xi_2a^2k^2$, $\sigma = 10 a$,  $ka=0.5$, $\lambda = 0.05g$.}
\end{figure}

\section{Classical light limit}
\label{sec:class_lim}

The case of a coherent state of electromagnetic field with the large photonic number $n$ is especially interesting for us. In this case one can neglect the quantum nature of the light and consider  electromagnetic field as a classical one replacing the electric field operator with its expectation value  $\bm{E} (x,t) = Re\{\mathbf{\mathfrak{E}} \exp[i(kx-\omega t)]\}$. In doing this we neglect any changes of the field state. Omitting  the local-field effects as before, we can write the Hamiltonian of the system "QD-chain - electromagnetic field" in the form $\hat {H}=\hat {H}_0+\hat {H}_T$, where the term
\begin{equation}
\label{h_0}
\hat {H}_0 =
    \frac{\hbar \omega _0}{2} \sum\limits_n {\hat {\sigma }_{zn}}-
    \frac{\hbar\Omega _R}{2}\sum\limits_n \left[\hat {\sigma }_n^+
    e^{i(nka-\omega t)}+\mathrm{H.c.}\right]
\end{equation}
describes Rabi oscillations in non-interacting QDs and $\Omega_R =  \bm{\mu}\mathbf{\mathfrak{E}}/\hbar$ is the Rabi frequency \cite{Scully}, The interdot interaction mechanism is independent on light properties, so the term $\hat {H}_T$ is defined as before by equation \eqref{h_t}.

The state vector of the system has the form:
\begin{equation}
\label{state_vector_cl}
\left| {\Psi (t)}
\right\rangle =\sum_p \left(A_p(t) \left|a_p \right\rangle +B_p(t)
\left| b_p \right\rangle \right),
\end{equation}
where $A_p(t)$, $B_p(t)$ are unknown functions. Now the equations of motion for them can now be obtained using approach of Sec.\ref{sec:mot_eq}. Introducing the decay factor $\lambda$  into the nonstationary Schr\"odinger equation and making continuous limit transition in the QD-chain similar to the Sec.\ref{sec:cont_lim},  we obtain the system of equations as follows:
\begin{equation}
\label{system_cl}
\begin{split}
\partial_t \bm{\Psi} = \left[2i\hat{\xi}-\frac{i\omega_0}{2}\hat{\sigma}_z-\frac{{\lambda}}{2}\hat{I}+\frac{i\Omega_R}{2}\,\hat{\kappa}(kx-\omega t)\right]\bm{\Psi}\\+ia^2\hat{\xi}\partial^2_x\bm{\Psi},
\end{split}
\end{equation}
where $\bm{\Psi}(x,t) = \left[ \begin{array}{c} A(x,t) \\ B(x,t) \end{array} \right]$.  Note, that spontaneous emission can be taken into account in the same way as in \ref{sec:mot_eq}, i.e. by describing it as interaction of QDs  with photon bath. In this case the damping constant $\lambda$ in Eq.\eqref{state_vector_cl} will be the sum of two partial constants. The system \eqref{state_vector_cl} can be considered as a particular case  of the general system  \eqref{system} and can be solved in the same way. The solution reads

\begin{equation}
\label{sol_cl}
\begin{split}
\bm{\Psi}(x,t)=&e^{i(kx-\omega t)\hat{\sigma}_z/2}e^{-\lambda t/2}\\
\times&\int\limits_{-\infty }^\infty {\hat{\rho}(h,t)\widetilde{\bm{\Phi}}(h,0) e^{i\{hx+[\vartheta_1(h)+\vartheta_2(h)]t/2\}}} dh.
\end{split}
\end{equation}
where the values  $\hat{\rho}(h,t)$, $\varphi^\pm (h,t)$, $\psi_h (t)$, $\delta_{\pm}(h)$, $\Omega(h)$, $\Delta_{eff}(h)$ are determined, respectively, by equalities \eqref{rho}, \eqref{phi}, \eqref{psi},  \eqref{delta}, \eqref{omega}, \eqref{delta_eff} if one makes the substitution  $2g\sqrt{n+1}\rightarrow\Omega_R$ (in doing that the photonic number dependence in $\hat{\rho}(h,t)$, $\varphi^\pm $, $\psi_h$, $\Omega$ disappears). The vector function $\widetilde{\bm{\Phi}}(h,0)$  is
determined by the initial conditions similar to Eq.\eqref{inv_f_transform_2}.

\begin{figure}[ht]
\includegraphics[width=0.48\textwidth,trim=35 15 28 25,clip]{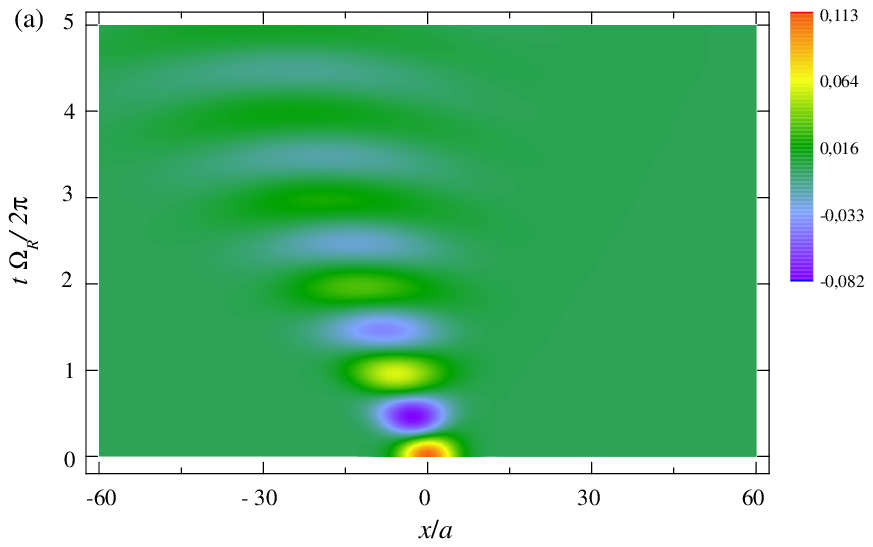}
\includegraphics[width=0.48\textwidth,trim=35 15 28 25,clip]{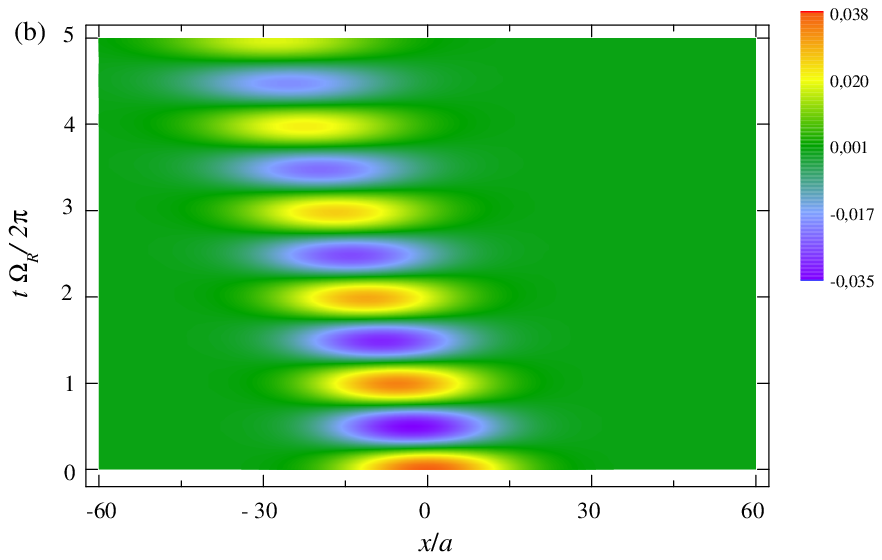}
\caption{\label{fig2} Space-time distribution of the inversion in
the QD in the classical light chain for a single Gaussian wavepacket $A(x,0)
=(1/\sqrt[4]{\pi\sigma^2})$ $ \exp(-x^2/2\sigma^2)$, $B(x,0)=0$, $\xi_1=3\Omega_R$, $\xi_2=0.9\xi_1$,
$\Delta = 2(\xi_1-\xi_2)+\xi_2a^2k^2$, $ka=0.33$, $2\pi\lambda/\Omega_R = 0.1$. a) $\sigma = 5 a$; b) $\sigma = 15 a$}
\end{figure}

\begin{figure}[htb]
\includegraphics[width=0.48\textwidth,trim=15 17 15 10,clip]{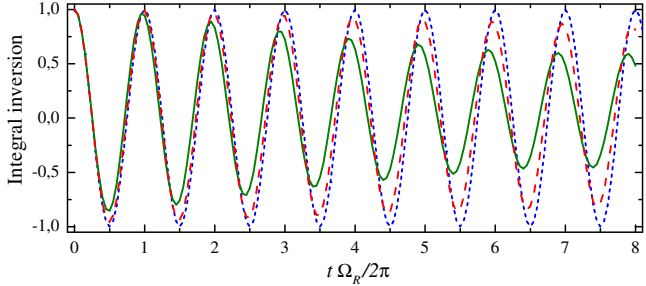}
\caption{\label{fig3} Temporal dependence of the  integral inversion in the classical light
for the input parameters as follows:  $\Delta = 0$, $k = 0$, $\xi_1 = \xi_2 = 3 \Omega_R$ (dotted line); $\Delta = \xi_2a^2k^2$, $ka = 0.33$, $\xi_1 = \xi_2 = 3 \Omega_R$ (solid line); $\Delta = 2(\xi_1-\xi_2)+\xi_2a^2k^2$, $ka = 0.33$, $\xi_1 =  3 \Omega_R$, $\xi_2 =  1.5 \Omega_R$ (dashed line). In all cases $\sigma = 5a$, $\lambda = 0$. }
\end{figure}

A typical space-time distribution of the inversion density $w(x,t)=a[|A(x,t)|^2-|B(x,t)|^2]$  is shown in fig.~\ref{fig2}.
As is seen from the figure, the space-time dynamics of Rabi wavepacket is similar to the one for the case of vacuum oscillations (but it is of substantially different physical nature: electromagnetic field is in coherent state with photon number $n\longrightarrow\infty$ and this state does not change when the system oscillates between ground and excited states in contradiction to the case of vacuum Rabi oscillation). Since the photon distribution does not change, collapses and revivals are absent. However there is another phenomenon of similar nature. One can see in fig.~\ref{fig3}, where the plots of the integral inversion are presented, that the  oscillations of this quantity at $k \neq 0$,  $\xi_1\neq\xi_2$  decrease with time even at $\lambda=0$, whereas such a damping is absent at $k = 0$, $\xi_1=\xi_2$ and integral inversion oscillates harmonically in the range from -1 to 1 (dotted curve in fig.~\ref{fig3}).
Such a behaviour  is the consequence of dependence of the effective detuning $\Delta_{eff}(h)$ (and therefore the Rabi frequency  $\Omega(h)$) on $h$. But if  $k = 0$ and  $\xi_1=\xi_2$, then $\Delta_{eff}=\Delta$ and $\Omega$ does not depend on $h$. The damping rate is controlled by values of wavenumber $k$ and coupling constants $\xi_{1,2}$. If the initial state of the system is the excited state ($B(x,0) = 0$), the damping rate is predominantly determined by product $\xi_2k$ (and damping is practically absent if $\xi_2k=0$), for the ground initial state ($A(x,0) = 0$) it is defined by $\xi_1k$. In the case of a mixed initial state all three quantity are important. 

Note, that more narrow wavepacket spreads faster, than the wider one (compare fig.~\ref{fig2}a and fig.~\ref{fig2}b).

In the weak coupling limit the indicated dephasing mechanism is analogous to the Landau damping in plasma.
\begin{figure}[htb]
\includegraphics[width=0.48\textwidth,trim=35 15 28 25,clip]{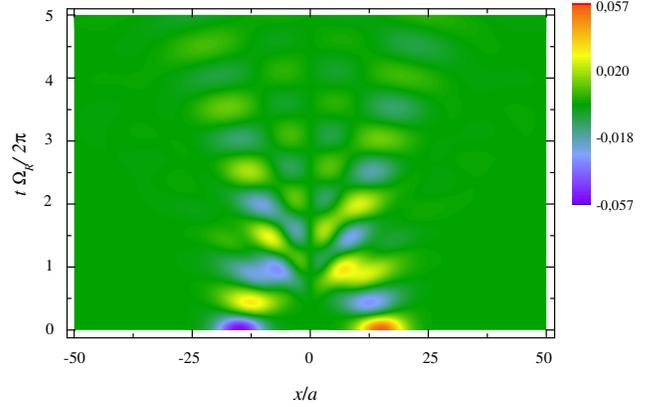}
\caption{\label{fig4} Space-time distribution of the inversion density in
the QD chain in the classical light for two counterpropagating identical Gaussian
wavepackets: $A(x,0) = \exp[-(x-3\sigma)^2/2\sigma^2]/\sqrt[4]{4\pi\sigma^2}$, $B(x,0) =
\exp[-(x+3\sigma)^2/2\sigma^2]/\sqrt[4]{4\pi\sigma^2}$, $\Delta = 0$, $ka=0.33$, $\sigma = 5 a$,  $\xi_1=\xi_2 = 3 \Omega_R$, $2\pi\lambda/\Omega_R = 0.1$.}
\end{figure}

Interaction  of two counterpropagating identical Gaussian Rabi
wavepackets colliding at  $x=0$ is shown in fig.~\ref{fig4}.
Unlike the case of quantum field, the space-time distribution of the inversion density is fully symmetrical with respect to the line  $x=0$ and   integral inversion does not oscillate: this quantity equals zero for all $t\geq 0$ and arbitrary values of $\Omega_R$.

\section{Influence of the local-field effects}
\label{sec:local_field}

Local fields are depolarization fields arising inside QDs due to the dipole-dipole electron-hole interaction. As a result, the field acting on the e-h pair  differs from mean field \cite{Slepyan_02}. The local-field action is described by Eq.\eqref{Deltah} and leads to nonlinear terms in the equations of motion (last terms in Eqs. \eqref{system_diskr_a}-\eqref{system_diskr_b}).  The presence of nonlinearity results in the qualitatively new features of Rabi oscillations in single QDs \cite{Slepyan_04,Slepyan_07, Paspalakis_06}.  The influence of the local-fields  on Rabi waves is the subject for future investigations. In this section we consider some aspects of this influence only and discuss some sufficient conditions of the  local-field  negligibility.

Let us consider finite (of the length $L$) QD-chain interacting with classical light. The relaxation processes are neglected (i.e. $\lambda = 0$). The equations of motion follow from  Eqs.\eqref{system_diskr_a}-\eqref{system_diskr_b} in the same way as Eq.\eqref{system} reduce to Eq.\eqref{system_cl} and have the form

\begin{align}
\label{system_1a}
\nonumber\partial_t A =\displaystyle{-\frac{i}{2}(\omega_0-4\xi)}A+\frac{i\Omega_R }{2}&B e^{i(kx-\omega t)}+\\
 i\xi& a^2\partial_x^2A -i\Delta\omega|B|^2A\,,
    \\ \rule{0in}{3ex}
    \label{system_1b}
\nonumber \partial_t B = \displaystyle{\frac{i}{2}(\omega_0+4\xi)}B+\frac{i\Omega_R}{2}A&e^{-i(kx-\omega t)}+\\
i\xi& a^2\partial_x^2B-i\Delta\omega|A|^2B\,.
\end{align}

We use the periodic boundary conditions (Born-von Karman conditions): 
\begin{equation}
    \label{born_karman}
A(L/2,t)=A(-L/2,t), \, B(L/2,t) = B(-L/2,t).
\end{equation}

Suppose that electromagnetic field are also satisfy periodic conditions, i.e. $kL=2\pi m$, where $m$ is integer number. Then the problem under consideration has a clear physical interpretation:  we are dealing with ringed QD-chain of the radius $R=L/2\pi$. QD-chain interacts with electromagnetic mode of whispering gallery type, which is excited in microcavity with rotation axis ( $e^{ikx} \rightarrow e^{im\varphi}$, $x \rightarrow R\varphi$, $\varphi$ is the azimuthal coordinate).

Let us seek partial solution of the system \eqref{system_1a}-\eqref{system_1b} in the form of traveling wave: $A(x,t) = u_0 e^{i(h+k/2)x}e^{-i(\nu+\omega/2)t}$, $B(x,t) = v_0 e^{i(h-k/2)x}e^{-i(\nu-\omega/2)t}$, where $u_0$, $v_0$ are constant amplitudes satisfying the normalization condition for wavefunction, $h$ and $\nu$ are real values. Ssubstituting $A$ and $B$ into Eqs.\eqref{system_1a}-\eqref{system_1b} we obtain
\begin{align}
\label{char_eq_2a}
& \left[\nu+\phi_1-\Delta\omega|v_0|^2\right]u_0 + \frac{\Omega_R}{2}v_0=0,  \\ \rule{0in}{4ex}
\label{char_eq_2b}
& \left[\nu+\phi_2-\Delta\omega|u_0|^2\right]v_0 + \frac{\Omega_R}{2}u_0=0,
\end{align}
where $\phi_{1,2}(h) = 2\xi_{1,2} \mp\Delta/2-\xi_{1,2}a^2(h\pm k/2)^2$. From boundary conditions \eqref{born_karman} one can obtain the quantization condition for  $h$: $h = \pi n/L$, $n$ is integer. Expressing $v_0$ from Eq.\eqref{char_eq_2b} and substituting into Eq.\eqref{char_eq_2a}, we obtain 
\begin{equation}
\label{eq_for_nu}
   \nu+\phi_1(h)- \frac{\Omega_R^2(\nu+\phi_2(h))}{4(\nu+\phi_2(h)-\Delta\omega|u_0|^2)^2}=0.
\end{equation}
The wavefunction normalization condition results in

\begin{equation}
\label{eq_u_0}
   \left(1+\frac{\Omega_R^2}{4|\nu+\phi_2(h)-\Delta\omega|u_0|^2|^2}\right)|u_0|^2=\frac{1}{L}.
\end{equation}

Eqs.\eqref{eq_for_nu}-\eqref{eq_u_0} form the closed system of equations with respect to  $\nu$ and $|u_0|^2$, which defines the dispersion law  of Rabi waves $\nu=\nu(h)=\nu(n)$ with regard to the local-field effects. It should be noted that only solutions with real-valued $\nu$ have physical meaning (otherwise the system \eqref{char_eq_2a}-\eqref{char_eq_2b} contradicts to the initial equations  \eqref{system_1a}-\eqref{system_1b}). For $L \rightarrow \infty$, $\Delta\omega \rightarrow 0$ the system \eqref{eq_for_nu}-\eqref{eq_u_0} reduces to the dispersion equation \cite{Slepyan_09} for Rabi waves in the infinite QD-chain interacting with classical light.

\begin{figure}[htb]
\includegraphics[width=0.48\textwidth,trim=12 15 15 5,clip]{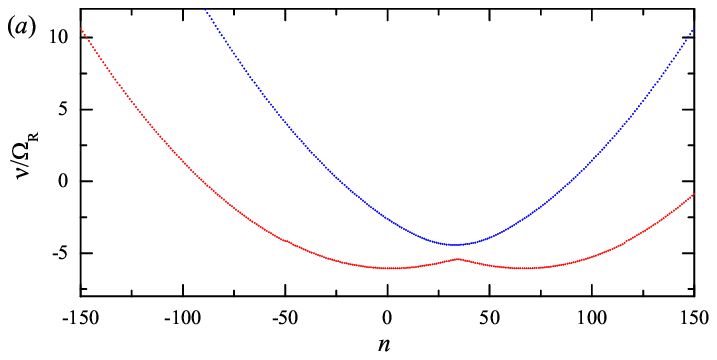}
\includegraphics[width=0.48\textwidth,trim=15 15 15 5,clip]{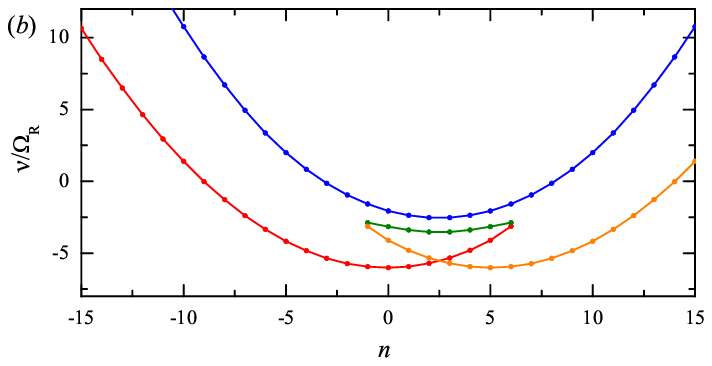}
\includegraphics[width=0.48\textwidth,trim=15 16 10 5,clip]{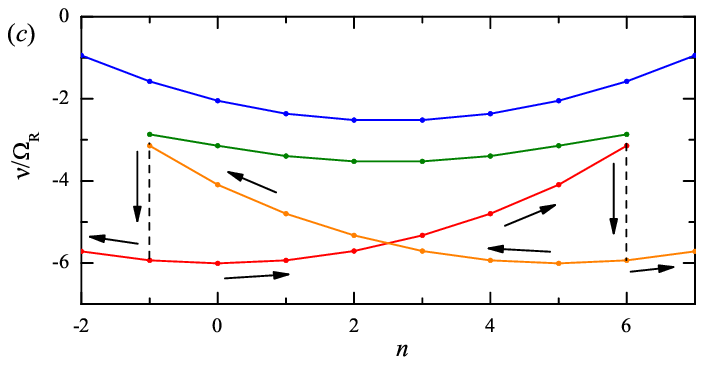}
\caption{\label{nonlin_disp} Dispersion curves for Rabi waves subject to local-field effects. a) $L = 200 a$, b), c) $L = 20 a$. In all cases $\xi_1 = \xi_2 = 3 \Omega_R$,  $ka=0.25\pi$, $\Delta = 0$, $\Delta\omega = 100\Omega_R$. }
\end{figure}

The dispersion characteristics of Rabi waves for two essentially different values of $L$ but identical $\Delta\omega$ and other parameters are shown in Fig.\ref{nonlin_disp}. The curves depicted  in Fig.\ref{nonlin_disp}a corresponds to the QD-chain of large radius. The nonlinear effects in this case are negligibly small. Dispersion curves do not feel nonlinear effects and are practically identical to obtained in \cite{Slepyan_09}. For QD-chain of small radius the situation becomes principally different due to nonlinear effects. The \textit{closed loop} appears in the  vicinity of any cross point and the number of the modes increases to four (Fig.\ref{nonlin_disp}b). The size of this loop decreases with decreasing the ratio $\Delta\omega/\Omega_R$.

The detailed structure of closed loop  is shown in Fig.\ref{nonlin_disp}c. When $h$ (and, correspondingly, $n$) changes adiabatically slow the motion along the loop takes place (it is shown by arrows in Fig.\ref{nonlin_disp}c).  The transitions from one branch to another are followed by simultaneous changes of Rabi frequency and Rabi wave group velocity. The trajectories of motion in different directions proves to be nonidentical. Thus, the loop has an hysteresis character. Interaction of the local-fields and Rabi waves leads to effect of bistability of quantum states in the QD-chain. The detailed analysis of different dispersion branches stability is the subject for future investigations.

 Let us now discuss the conditions of the local field effects neglecting in equations \eqref{system_diskr_a}--\eqref{system_diskr_b}. Note, that for $\xi_{1,2}\rightarrow0$ the system \eqref{system_diskr_a}--\eqref{system_diskr_b} transforms to equations (24) in \cite{Slepyan_07}, where Rabi oscillations in single QD are described with account of local field effects. An analysis of  equations (24) in \cite{Slepyan_07} has shown that local field effects are negligibly small for the case of sufficiently large Rabi frequencies, namely, $\Omega_{\left\langle n\right\rangle}\gtrsim \Delta\omega$, where $\Omega_{\left\langle n\right\rangle} = \sqrt{\Delta^2+4g^2(\left\langle n\right\rangle+1)}$, $\left\langle n\right\rangle$  is average photon number. If this condition is fulfilled for coherent state, the conventional picture of Rabi oscillations (the alternation of collapses and revivals) takes place. In the opposite case qualitatively new regime of oscillations arises. This regime is due to the local field effects and is described in detail in\cite{Slepyan_07}. In this work we are interested only in first regime. In our case the role of $\Omega_{\left\langle n\right\rangle}$ is played by frequencies $\nu_{1,2}(n,h)$, which are determined by Eq.\eqref{disp_law}. Therefore it is reasonable to suggest that for traveling Rabi waves the local fields effects are negligible under the condition 
 \begin{equation}
 \label{loc_f_neg}
 \nu_{1,2}(n,h)\gtrsim \Delta\omega, 
 \end{equation}
 where $\Delta\omega$ is determined by Eq.\eqref{dep_shift}. 

In the case of Rabi-wavepackets  general line of reasoning is the same as for Eq.\eqref{loc_f_neg} and leads to the following condition of the local-field negligibility: $\nu_{1,2}(\left\langle n\right\rangle,~h^{(0)}_{1,2})\gtrsim \Delta \omega$.

\section{Correlation functions}
\label{sec:correlators}

\begin{figure}[htpb]
\includegraphics[width=0.48\textwidth,trim=25 10 10 10,clip]{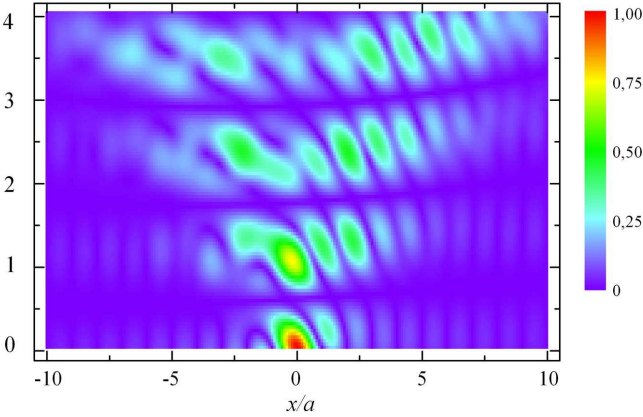}
\caption{\label{correl} The space-time dependence of  normalized correlation function $g^{(1)}(x,0,t,0)$ for light in Fock state with $n = 5$ and the input parameters as follows: $\xi_1 =\xi_2 = 10 g$,  $\Delta = \xi_2 k^2 a^2$, $\sigma = 5a$, $A_n(x,0) =  \exp(-x^2/2\sigma^2)/\sqrt[4]{\pi\sigma^2}$, $B_{n+1}(x,0) = 0$.}
\end{figure}

In this section we will calculate exciton-exciton and exciton-photon multi-time correlation functions. For this purpose it is convenient to pass to the Heisenberg picture and to use the discrete model of QD-chain (transition to the continuous limit will be made on the final stage). Thus,  we should look for a time dependence of operators $\hat{\sigma}_{p}^{\pm}$, i.e. we shall seen to find the evolution operator. It can be obtained from  \eqref{system_diskr2} (for details see Appendix \ref{sec:evol_operator}) and has the form
\begin{equation}
\label{ev_operator}
\begin{split}
&\hat{U}(t,0) = \frac{a}{2\pi}\sum\limits_n\sum\limits_{p,q}\int\limits_{-\pi/a}^{\pi/a} \left\{{\varphi}_n^- (h,t) e^{i\delta_h^+t} \left|a_p,n\right\rangle\left\langle a_q,n\right|\right.+\\
&{\psi}_n (h,t)\left[e^{i\delta_h^+t}\left|a_p,n\right\rangle\left\langle b_q,n+1\right|+e^{i\delta_h^-t}\left|b_p,n+1\right\rangle\left\langle a_q,n\right|\right]\\
&\left.+{\varphi}_n^+ (h,t) e^{i\delta_h^-t} \left|b_p,n+1\right\rangle\left\langle b_q,n+1\right| \right\}e^{ih(p-q)a}dh.
\end{split}
\end{equation}
Here ${\varphi}_n ^\pm (h,t)$, ${\psi}_n (h,t)$, ${\Omega}_n(h)$ are given by formulas  \eqref{phi}, \eqref{psi}, \eqref{omega}. Quantities $\delta_h^{\pm}$,  $\Delta_{eff}(h)$ are defined by expressions \eqref{delta},  \eqref{delta_eff} if one substitutes $\vartheta_{1,2} \rightarrow 2\xi_{1,2} \cos[(h\pm k/2)a]$. Note, that if one makes approximation $\cos[(h\pm k/2)a] \cong1-(h\pm k/2)^2a^2/2$ (this is the usability condition for continuous limit), the  quantities $\delta_h^{\pm}$,  $\Delta_{eff}(h)$ coincide with expressions \eqref{delta},  \eqref{delta_eff}.

From expression \eqref{ev_operator} we obtain by using the formula $\hat{\sigma}_p^{\pm}(t)=\hat{U}^+(t,0)\hat{\sigma}_p^{\pm}\hat{U}(t,0)$ the exciton-exciton correlation functions $G^{(1)}_{p,q}(t,t')\equiv\left\langle \hat{\sigma}_{p}^+(t)\hat{\sigma}_{q}^-(t')\right\rangle$  as follows:
\begin{widetext}
\begin{gather}
\label{e-e_corr_diskr}
\nonumber G^{(1)}_{p,q}(t,t')= \frac{a^4 }{16\pi^4} \sum\limits_n\sum\limits_{l,r} \iiint\limits_{-\pi/a}^{ \ \ \ \pi/a} e^{ih'(p-q)a}e^{i(qg-ph)a}e^{i(lh-rg)a}e^{-i\Delta (t-t')} \left[\varphi_n^+ (h',t)\varphi_n^-(h',t')-\psi_n(h',t)\psi_n(h',t')\right]\\
[\varphi_{n+1}^+(h,t)\varphi_{n+1}^-(g,t')u_{l,n+1}^*(0)u_{r,n+1}(0)+
\psi_{n+1}(h,t)\psi_{n+1}(g,t')v_{l,n+2}^*(0)v_{r,n+2}(0) \\
\rule{0in}{3ex} \nonumber+\varphi_{n+1}^+(h,t)\psi_{n+1}(g,t')u_{l,n+1}^*(0)v_{r,n+2}(0)-
\psi_{n+1}(h,t)\varphi_{n+1}^-(g,t')v_{l,n+2}^*(0)u_{r,n+1}(0)]dh'dhdg,
\end{gather}
 \end{widetext}
where $u_{p,n}(t)=A_{p,n}(t)e^{-i(kpa-\omega_0 t)/2}$, $v_{p,n+1}(t)=B_{p,n+1}(t)e^{i(kpa-\omega_0 t)/2}$. The continuous limit transition can be made  by means of the following substitutions: $pa~\rightarrow~x$, $qa~\rightarrow~x'$, 
$\sum\limits_l\rightarrow\frac{1}{a}\int\limits_{-\infty}^{ \infty}...\,dz$, $\sum\limits_r\rightarrow\frac{1}{a}\int\limits_{-\infty}^{ \infty}...\,dz'$.

One can see, that in general case $G^{(1)}_{p,q}(t,t')$ can not be represented as $G^{(1)}_{p-q}(t,t')$ or $G^{(1)}_{p,q}(t-t')$. It indicates the absence both spatial and time homogeneity.

Let us consider case of $B_{p,n+1}(0)=0$ and $t'=0$. Then $\psi_{n}(h,t')=0$, $\varphi_n^{\pm}(h,t')=1$ and taking  into account that $\int_{-\pi/a}^{\pi/a}e^{ig(q-r)a}dg = 2\pi\delta_{qr}/a$, we obtain rather simple expression for correlation function:

\begin{align}
\nonumber &G^{(1)}_{p,q}(t,0)=\frac{a^3 e^{-i\Delta t}}{8\pi^3} \sum\limits_n\sum\limits_{l} \iint\limits_{-\pi/a}^{ \ \ \ \pi/a} e^{ih'(p-q)a}e^{ih(l-p)a} \\
&\varphi_n^+ (h',t)\varphi_{n+1}^+(h,t)u^*_{l,n+1}(0)u_{q,n+1}(0)dh'dh,
\end{align}
or, in continuous limit,
\begin{align}
\nonumber &G^{(1)}(x,x',t,0)=\frac{a^2 e^{-i\Delta t}}{8\pi^3} \sum\limits_n\int\limits_{-\infty}^{\infty} \iint\limits_{-\pi/a}^{ \ \ \ \pi/a} e^{ih'(x-x')}e^{ih(z-x)} \\
&\varphi_n^+ (h',t)\varphi_{n+1}^+(h,t)u^*_{n+1}(z,0)u_{n+1}(x',0)dh'dhdz.
\end{align}

Plot of normalized correlation function
\begin{equation}
g^{(1)}(x,x',t,t')=\frac{G^{(1)}(x,x',t,t')}{\sqrt{G^{(1)}(x,x,0,0)G^{(1)}(x',x',0,0)}}
\end{equation}
for $x'=0$, $t'=0$ is presented on Fig.\ref{correl}. The external field  is supposed to be in a single-photon initial state and the electron in QD-chain is in the excited initial state ($B_{n+1}(x,0)=0$).

The correlation function of the polarization operator $\hat{\sigma}_p^+(t)$ with the operator of the external field $\hat{E}_p^-(t)=\mathcal{E}\hat{a}e^{i(kpa-\omega t)}$ (exciton-photon correlators) appears as

\begin{widetext}
\begin{gather}
\left\langle \hat{E}_q^-(t')\hat{\sigma}_p^+(t)\right\rangle= \frac{a^2 \mathcal{E}}{4\pi^2} \sum\limits_n\sum\limits_{l,r}\iint\limits_{-\pi/a}^{ \ \ \ \pi/a} dhdh' \{\varphi_{n+1}^+ (h',t)\sqrt{n+1}\left[\psi_n (h,t)u^*_{r,n}(0)u_{l,n}(0)+\varphi_n^{+}(h,t)u^*_{r,n}(0)v_{l,n+1}(0)\right] \\ \nonumber -\psi_{n+1}(h',t)\sqrt{n+2}\left[\psi_n(h,t)v^*_{r,n+1}(0)u_{l,n}(0)+\varphi_n^+(h,t)v^*_{r,n+1}(0)v_{l,n+1}(0)\right] \} e^{ipa(h-h')}e^{i(h'r-hl)a}e^{i(kpa-\omega_0 t)}e^{i\omega(t-t')}.
\end{gather}
\end{widetext}

\section{Conclusion}
\label{sec:conclusion}

 The main results, obtained in this work, can be summarized as follows:

i) In QD-chain with tunneling coupling in strong electron-photon coupling regime the space propagation of Rabi oscillations (Rabi waves) takes place. For  propagation of Rabi waves the wave vector of the photon mode must have a nonzero component along the chain.  Characteristics of the Rabi waves  depend strongly on relations between parameter of electron-photon coupling, frequency deviation and transparency factors of tunneling barriers for both of levels. 

ii) Traveling  Rabi waves are  quantum states of  the QD-chain dressed by radiation. The
qualitative distinction of this states from states of single dressed atom \cite{Cohen-Tannoudji_1998} is the space-time modulation
of dressing parameter  by traveling wave law. Traveling  Rabi waves can be interpreted as entangled states of e-h pair and photons. These states are characterized by the dependence of energy on quasi-momentum and can be treated as new type of quasiparticles (in this paper they are called rabitons).

iii) There are two traveling Rabi modes with different frequencies of Rabi oscillations at a given value of wavenumber. The range of  Rabi oscillations frequencies is limited by the critical value, which is different for both modes. The QD-chain is opaque for Rabi modes with frequencies lesser than critical one. The critical frequencies as well as dispersion characteristics depend on number of photons in the mode.

iv) In general case the propagation of Rabi wavepackets occurs  in the form of four partial packets, which are characterized by different amplitudes and velocities of motion. Rabi wavepackets transfer  energy, inversion, quasimomentum,  electron-electron and electron-photon quantum correlations along the chain. Particularly, in the case of QD-chain interacting with quantum light in the coherent state the known collapses-revivals picture \cite{Scully} is drastically modified due to propagation effect: collapses and revivals take place in different points of space.

v) One pair of Rabi wavepackets exists due to the ground state contribution in initial state of the system, another one appears  due to the contribution of excited state (generally, initial state is arbitrary superposition of ground and excited states). Each pair of packets moves through its own potential barrier. If one of the barriers is completely opaque, then corresponding pair of wavepackets does not move and does not spread. 

vi) Rabi wavepackets propagation along the QD-chain is followed by the transformation of quantum
light statistics (for example, in initially coherent light incoherent component appears). If specific conditions are fulfilled then incoherent part can be decreased to zero.

vii) In particular cases a number of Rabi wavepackets can be diminished. There are two mechanisms of such diminution: tending to zero  of  partial wavepacket amplitude and merging of the packets due to their velocities equality.

Rabi waves can take place in a number of other distributed systems strongly coupled with electromagnetic field.

The example are superconducting circuits based on Josephson junctions, which are currently the most experimentally advanced solid-state qubits \cite{Blais_07}. It is evident, for example, that the chain of qubits placed inside a high-Q transmission-line resonator due to the qubit-qubit capacitance coupling will support the Rabi waves propagation similar to described in this article.

Rabi waves effect can be practically used in nanoelectronics, quantum computing, quantum informatics. 

\section*{Acknowledgments}
The work of G. Ya. Slepyan was partially carried out
during the stay at the Institut f\"{u}r Festk\"{o}rperphysik, TU
Berlin, and supported by the Deutsche Forschungsgemeinschaft (DFG).
Authors are grateful to Dr. S. A. Maksimenko, Dr. D. S. Mogilevtsev, Dr. J. Haverkort and  Dr. A. M. Nemilentsau for stimulative discussions.

\appendix

\section{Evolution operator}
\label{sec:evol_operator}

In this appendix, the evolution operator for a discrete QD-chain (Eq.\eqref{ev_operator}) is obtained from Eq.\eqref{system_diskr2}  Let us introduce new variables $u_{p,n}=A_{p,n}e^{-i(kpa-\omega_0 t)/2}$, $v_{p,n+1}=B_{p,n+1}e^{i(kpa-\omega_0 t)/2}$. It leads to the system 

\begin{align}
\label{system_diskr3_a}
\nonumber \partial_tu_{p,n}&=i\xi_1\left(u_{p-1,n}e^{-ika/2}+u_{p+1,n}e^{ika/2}\right)\\
&-i g\sqrt{n+1}v_{p,n+1} e^{i\Delta t},\\
\label{system_diskr3_b}
\nonumber\partial_tv_{p,n+1}&=i\xi_2\left(v_{p-1,n+1}e^{ika/2}+v_{p+1,n+1}e^{-ika/2}\right)\\
&-i g\sqrt{n+1}u_{p,n} e^{-i\Delta t}.
\end{align}
Coefficients in this system do not depend from QD-number $p$. It allows us to  seek the solution in the form $u_{p\pm1,n}=u_{p,n}e^{\pm iha}$, $v_{p\pm1,n+1}=v_{p,n+1}e^{\pm iha}$, $ha\in[-\pi,\pi]$. Then the equations for different QDs become independent and may be written as 
\begin{align}
\label{system_diskr4_a}
&\partial_tu_{p,n}=2i\xi_1 \cos [(h+k/2)a] u_{p,n}-i g\sqrt{n+1}v_{p,n+1} e^{i\Delta t},\\
\label{system_diskr4_b}
&\partial_tv_{p,n+1}=2i\xi_2 \cos [(h-k/2)a]v_{p,n+1}-i g\sqrt{n+1}u_{p,n} e^{-i\Delta t}.
\end{align}
Expressing $v_{p,n+1}$ from \eqref{system_diskr4_a} and substituting into \eqref{system_diskr4_b}, one obtains the set of independent second-order ordinary differential equations for the variables $u_{p,n}$
\begin{equation}
\label{eq_for_u}
\frac{d^2u_{p,n}}{dt^2}-i \delta_h \frac{d u_{p,n}}{dt}+\frac{\Omega_n^2(h)-\delta_h^2}{4} u_{p,n}=0,
\end{equation}
where $\delta_h =\Delta+2\xi_1 \cos [(h+k/2)a]+2\xi_2 \cos [(h-k/2)a]$.
The solution of \eqref{eq_for_u} reads 
\begin{equation}
\label{u_p}
u_{p,n}(t)=C_1^n(0)e^{i\left[\delta_h+\Omega_n(h)\right]t/2}+C_2^n(0)e^{i\left[\delta_h-\Omega_n(h)\right]t/2}
\end{equation}
with the constants  $C_{1,2}^n(0)$ to be determined from initial conditions. Writing the latter in the form $u_{p,n}(0)=a_ne^{ihpa}$, $v_{p,n+1}(0)=b_{n+1}e^{ihpa}$,  we rewrite \eqref{u_p} as
\begin{equation}
u_{p,n}(t)=\left[\alpha_ne^{i\left[\delta_h^+-\Omega_n(h)/2\right]t}+\beta_ne^{i\left[\delta_h^++\Omega_n(h)/2\right]t}\right] e^{ihpa}
\end{equation}
and
\begin{equation}
v_{p,n+1}(t)=\left[\gamma_ne^{-i\left[\delta_h^-+\Omega_n(h)/2\right]t}+\delta_ne^{-i\left[\delta_h^--\Omega_n(h)/2\right]t}\right] e^{ihpa},
\end{equation}
where
\begin{align}
\alpha_n&=\frac{\Omega_n(h)+\Delta_h}{2\Omega_n(h)}a_{n}+\frac{g\sqrt{n+1}}{\Omega_n(h)}b_{n+1},\\
\beta_n&=\frac{\Omega_n(h)-\Delta_h}{2\Omega_n(h)}a_{n}-\frac{g\sqrt{n+1}}{\Omega_n(h)}b_{n+1},\\
\gamma_n&=\frac{g\sqrt{n+1}}{\Omega_n(h)}a_{n}+\frac{\Omega_n(h)-\Delta_h}{2\Omega_n(h)}b_{n+1},\\
\delta_n&=-\frac{g\sqrt{n+1}}{\Omega_n(h)}a_{n}+\frac{\Omega_n(h)+\Delta_h}{2\Omega_n(h)}b_{n+1}.
\end{align}

Basis wavefunction has the form

\begin{equation}
\left| {\Psi_h (t)} \right\rangle =\sum\limits_n \sum\limits_p \left(u_{p,n}(t) \left|a_p,n \right\rangle +v_{p,n+1}(t) \left| b_p,n+1 \right\rangle \right) .
\end{equation}

We now can represent the required wavefunction as Fourier integral with basis functions $\Psi_h (t)$:
\begin{equation}
\label{wave_f}
\left| {\Psi (t)} \right\rangle =\int\limits_{-\pi/a}^{\pi/a}M(h)\left| {\Psi_h (t)} \right\rangle dh,
\end{equation}
where $M(h)$ is unknown weighting function. To determine it let us take into account, that

\begin{equation}
\left| {\Psi (0)} \right\rangle =\sum\limits_{n,p}\int\limits_{-\pi/a}^{\pi/a}M(h)[a_n\left| a_p,n \right\rangle +b_{n+1}\left| b_p,n+1 \right\rangle]e^{ihpa}dh,
\end{equation}
where
\begin{equation}
\left\langle a_p,n\right.\left| {\Psi (0)} \right\rangle =a_n\int\limits_{-\pi/a}^{\pi/a}M(h) e^{ihpa}dh 
\end{equation}
and the same for $\left\langle b_p,n+1\right.\left| {\Psi (0)} \right\rangle$.
Then employing inverse Fourier transform we have

\begin{equation}
\label{a_n}
a_n M(h) =\frac{a}{2\pi}\sum\limits_{q=-\infty}^{\infty}\left\langle a_q,n\right.\left| {\Psi (0)} \right\rangle e^{-ihqa}
\end{equation}
and analogously for $b_{n+1}$.

Substituting \eqref{a_n} into \eqref{wave_f}, we obtain $\left| {\Psi (t)} \right\rangle = \hat{U}(t,0)\left| {\Psi (0)} \right\rangle$, with  $\hat{U}$ given by the Eq.\eqref{ev_operator}. These calculations allow one to pass from Schr\"{o}dinger picture to the Heisenberg one. They are used in the calculations of different types of correlators.


\begin{thebibliography}{99}
\bibitem{ref01} In Hamiltonian $\hat{H_f}$ the constant term describing the photon vacuum contribution is omitted. In singlemode assumption its influence identically equal to zero. In so doing we are not interested in effects of spontaneous decay and Lamb shift. This effects are caused by interaction of the system with photon vacuum and can be taken into account in multimode assumption only.
\end{thebibliography}
\end{document}